\documentclass[structabstract]{aa}  
\usepackage{graphicx}
\usepackage{lscape}
\usepackage{amsmath}
\usepackage{txfonts}
\usepackage{url}
\usepackage{natbib}
\usepackage{xcolor}
\usepackage{color}

\begin{document}

\title{Circumstellar medium around rotating massive stars at solar metallicity}

\author{Cyril Georgy\inst{\ref{inst1},\ref{inst2}} \and Rolf Walder\inst{\ref{inst2}} \and Doris Folini\inst{\ref{inst2}} \and Andrei Bykov\inst{\ref{inst3}} \and Alexandre Marcowith\inst{\ref{inst4}} \and Jean M. Favre\inst{\ref{inst5}}}

\authorrunning{Georgy et al.}

\institute{Astrophysics group, EPSAM, Keele University, Lennard-Jones Labs, Keele, ST5 5BG, UK\\
                   \email{c.georgy@keele.ac.uk}\label{inst1} \and
                   \'Ecole Normale Sup\'erieure, Lyon, CRAL, UMR CNRS 5574, Universit\'e de Lyon, France\label{inst2} \and
                   Ioffe Physical Technical Institute of the Russian Academy of Sciences, Saint Petersburg, Russia\label{inst3} \and Laboratoire Univers et Particules de Montpellier, Universit\'e Montpellier 2, CNRS/IN2P3, CC 72,\\
Place Eug\`ene Bataillon, F- 34095 Montpellier Cedex 5, France\label{inst4} \and CSCS - Swiss National Supercomputing Centre, Via Trevano 131, 6900 Lugano, Switzerland\label{inst5}}

\date{Received ; accepted }

% RESUME %%%%%%%%%%%%%%%%%%%%%%%%%%%%%%%%%
\abstract
  % context heading (optional) leave it empty if necessary  
{}
  % aims heading (mandatory)
{Observations show nebulae around some massive stars but not around others. If observed, their chemical composition is far from homogeneous. Our goal is to put these observational features into the context of the evolution of massive stars and their circumstellar medium (CSM) and, more generally, to quantify the role of massive stars for the chemical and dynamical evolution of the ISM.}
  % methods heading (mandatory)
{Using the A-MAZE code, we perform 2d-axisymmetric hydrodynamical simulations of the evolution of the CSM, shaped by stellar winds, for a whole grid of massive stellar models from $15$ to $120\, M_{\sun}$ and following the stellar evolution from the zero-age main-sequence to the time of supernova explosion. In addition to the usual quantities, we also follow five chemical species: H, He, C, N, and O.}
  % results heading (mandatory)
{We show how various quantities evolve as a function of time: size of the bubble, position of the wind termination shock, chemical composition of the bubble, etc. The chemical composition of the bubble changes considerably compared to the initial composition, particularly during the red-supergiant (RSG) and Wolf-Rayet (WR) phases. In some extreme cases, the inner region of the bubble can be completely depleted in hydrogen and nitrogen, and is mainly composed of carbon, helium, and oxygen. We argue why the bubble typically expands at a lower rate than predicted by self-similarity theory. In particular, the size of the bubble is very sensitive to the density of the ISM, decreasing by a factor of $\sim 2.5$ for each additional dex in ISM density. The bubble size also decreases with the metallicity of the central star, because low-metallicity stars have weaker winds. Our models qualitatively fit the observations of WR ejecta nebulae.}
  % conclusions heading (optional), leave it empty if necessary 
{}


\maketitle
%%%%%%%%%%%%%%%%%%%%%%%%%%%%%%%%%%%%%%%%

% INTRODUCTION %%%%%%%%%%%%%%%%%%%%%%%%%%%%%%
\section{Introduction}

Massive stars are the main drivers of the chemical enrichment of and kinetic energy deposition in the interstellar medium (ISM), first through their winds and then during the explosion of the supernova (SN). Since the pioneering analytical and spherically symmetric work of \citet{Weaver1977a,Weaver1978a}, the development of even more efficient and refined hydrodynamical codes has allowed better and better understanding of the evolution of the circumstellar medium (CSM) surrounding massive stars. \citet{Garcia-Segura1996b,Garcia-Segura1996a} performed one of the first multi-D simulation of massive star environments, following different mass-loss events: main-sequence (MS) wind, red-supergiant (RSG) wind, Wolf-Rayet (WR) wind, etc. More recently, \citet{Freyer2003a,Freyer2006a} have studied the energetic impact of stellar winds on the ISM in more detail. The effect of stellar rotation has been discussed in various works \citep{Chita2007a,Chita2008a,vanMarle2008a}, as has the inclusion of ionisation in the computations \citep{Toala2011a,Dwarkadas2013a}. The chemical enrichment of the bubble created by the winds has also been studied by \citet{Kroger2006a}. More extensive studies of the evolution of the CSM around whole grids of stellar models have been done by \citet{Eldridge2006b}.

Recently, the Geneva stellar evolution group has released a new model grid at solar metallicity ($Z=0.014$), with their most recent code \citep{Ekstrom2012a}, including updated opacities and nuclear reaction rates, as well as the effect of internal rotation. These models were calibrated to reproduce a variety of observed characteristics: the characteristics of the Sun at its present age, the width of the MS in the Hertzsprung-Russell diagram (HRD), the position of red (super-) giants in the HRD, and the observed chemical composition and averaged surface velocities of B-type dwarf stars. They also succeed in partially reproducing the observed population of WR stars (and subtypes) at solar metallicity \citep{Georgy2012b}. The goals of this paper are to simulate wind blown bubbles for the whole set of rotating models of the Geneva grid and to discuss both common and specific characteristics. We address, in particular, the size, temperature, and density of the bubble, its chemical composition, and potential (non-) appearance as a nebula, all as a function of time and initial mass of the central star and with special attention to the effects of various successive wind episodes. We do not aim to perform a very-high resolution study of a peculiar case of stellar evolution, as is usually done \citep[see recently, \textit{e.g.},][]{vanMarle2012a}, but want to cover a wide variety of different cases instead.
 
The paper is organised as follows. Section~\ref{SecStellarModel} briefly recalls some important trends concerning the evolution of the central stars and explains the extraction of the wind properties (velocity, density) from the stellar evolution models. Section~\ref{SecHydroCode} presents the hydrodynamical code used in this work, as well as the set-up of the simulations. Our results are presented and discussed in Section~\ref{SecResults}, conclusions are drawn in Section~\ref{SecConclu}.

\section{The input model}\label{SecStellarModel}

\subsection{Stellar models}

We use as input the most recent grid of rotating stellar models provided by the Geneva group \citep{Ekstrom2012a}. We only consider the following massive star models (see Table~\ref{TabMassLoss}): $15$, $20$, $25$, $32$, $40$, $60$, $85$, and $120\, M_{\sun}$. Models of lower mass stars have radiative winds that are too weak to be of interest in this framework. The considered high-mass models were computed from the zero-age main sequence up to the end of central carbon burning and cover the main stages of stellar evolution: main sequence, RSG, or WR phase, if any. After carbon is exhausted in the centre of the star, the core evolution becomes so fast that the surface is insensitive to what happens in the core, and thus evolves no more. 

Among the available data, the following ultimately enter our simulations of the CSM: the luminosity and effective temperature of the star, its mass-loss rate, and its surface chemical composition, all of them being provided at each time step of the stellar evolution computation. The mass-loss rates are determined according to several prescriptions: \citet{Vink2001a}, if applicable, or \citet{deJager1988a} otherwise for stars hotter than $\log(T_\text{eff}/\text{K}) > 3.9$; \citet{deJager1988a} for red supergiants with $\log(T_\text{eff}/\text{K}) > 3.7$; \citet{Crowther2001a} when $\log(T_\text{eff}/\text{K}) < 3.7$; \citet{Nugis2000a} and \citet{Grafener2008a} for WR stars \citep[for more details, see][]{Ekstrom2012a}.

For the massive star models, we have the following trends for the evolution\citep{Georgy2012b}:
\begin{itemize}
\item The $15\, M_{\sun}$ follows a ``classical'' evolution, spending its MS in the blue part of the HRD, then crossing it up to the red supergiant branch, and ending its life on the red side of the HRD.
\item The $20$ and $25\, M_{\sun}$ models, after spending some time as a red supergiant, cross back the HRD up to $\log(T_\text{eff}/\text{K})\sim 4.3$, and enter a WR phase at the very end of their lives, as WNL stars.
\item The $32$ and $40\, M_{\sun}$ never reach the red supergiant branch, with a minimal $\log(T_\text{eff}/\text{K})\sim 3.8$. Then, they become WR stars, being successively WNL, WNE, and finally WC stars, with very high $\log(T_\text{eff}/\text{K}) \sim 5.2$.
\item Models more massive than $60\, M_{\sun}$ already become WR during the MS, completely avoiding the region of the HRD with $\log(T_\text{eff}/\text{K}) < 4$.
\end{itemize}

\begin{table}
\begin{center}
\caption{Grid of models and mass lost in each stellar evolution phase.}
\label{TabMassLoss}
\begin{tabular}{c|ccc||c|ccc}
$M_\text{ini}$ & MS & RSG & WR & $M_\text{ini}$ & MS & RSG & WR \\
$[M_{\sun}]$ &$[M_{\sun}]$ & $[M_{\sun}]$ & $[M_{\sun}]$ & $[M_{\sun}]$ & $[M_{\sun}]$ & $[M_{\sun}]$ & $[M_{\sun}]$\\
\hline
 \rule[0mm]{0mm}{3mm}$15$ & $0.3$ & $3.6$ & -- & $40$ & $8.0$ & $11.7$ & $8.0$\\
 $20$ & $0.5$ & $12.3$ & $0.02$ & $60$ & $15.6$\tablefootmark{1} & -- & $26.5$\\
 $25$ & $1.4$ & $13.5$ & $0.4$ & $85$ & $21.4$\tablefootmark{1} & -- & $37.2$\\
 $32$ & $3.9$ & $8.4$ & $9.6$ & $120$ & $32.5$\tablefootmark{1} & -- & $68.4$\\
\end{tabular}
\tablefoot{
\tablefoottext{1}{For these models already reaching the WR phase during the MS, the MS mass loss includes the amount of mass lost before the WR phase.}
}
\end{center}
\end{table}

A summary of the amount of mass lost during each stage is given in Table~\ref{TabMassLoss}.

\subsection{Extraction of the wind parameters: velocity and density}\label{SubSecExtraction}

In this section, we describe how we reconstruct the physical parameters of the stellar winds from the surface characteristics of the stellar models. First, the velocity $v_\infty$ of the wind far from the star is obtained using the relations by \citet{Kudritzki2000a}:
\begin{equation}
v_\infty = \left\lbrace\begin{array}{ll}
2.65 v_\text{esc} & T_\text{eff} \geq 21000\,\text{K} \\
1.4 v_\text{esc} & 10000\,\text{K} <T_\text{eff} < 21000\,\text{K} \\
 v_\text{esc} & T_\text{eff} \leq 10000\,\text{K},
\end{array}\right.
\end{equation}
with $v_\text{esc}$ the escape velocity at the stellar surface. To determine $v_\text{esc}$, we follow the procedure described in \citet[][see their eq.~16]{Lamers1996a}, and note that for hot massive stars it is crucial to account for the radiation pressure. The above formulation of $v_\infty$ is suitable for hot stars. For cool stars ($\log(T_\text{eff}/\text{K}) < 3.85$), we set the velocity to an average value for red supergiant winds of $25\,\text{km}\cdot\text{s}^{-1}$. 

The density of the wind at a distance $r$ (far enough) from the star of radius $R_\star$ is given by
\begin{equation}
\rho(r) = \frac{F_\text{m}(r)}{v_\infty},
\end{equation}
with $F_\text{m}(r)$ the mass flux (per unit time and surface). This mass flux is computed as a function of the flux at the surface with $F_\text{m}(r) = F_\text{m}(R_\star)\left(\frac{r}{R_\star}\right)^{-2}$. Even if the effect remains small for stars rotating far from their critical velocity, the anisotropy of the stellar wind is accounted for. The mass flux at the stellar surface is computed as in \citet{Georgy2011a}. For the initial velocity considered in this work, the maximal contrast between the polar and equatorial mass flux is $\sim 0.86$. We thus have $v_\infty(\theta)$ and $\rho(r,\theta)$ ($\theta$ being the colatitude) instead of a purely spherical symmetry. Upon mapping of the stellar wind parameters to radius $r$ angular momentum conservation of the wind is enforced.

\subsection{Typical mass-loss histories}\label{Sec_MassLossHistory}

\begin{figure}
\begin{center}
\includegraphics[width=.5\textwidth]{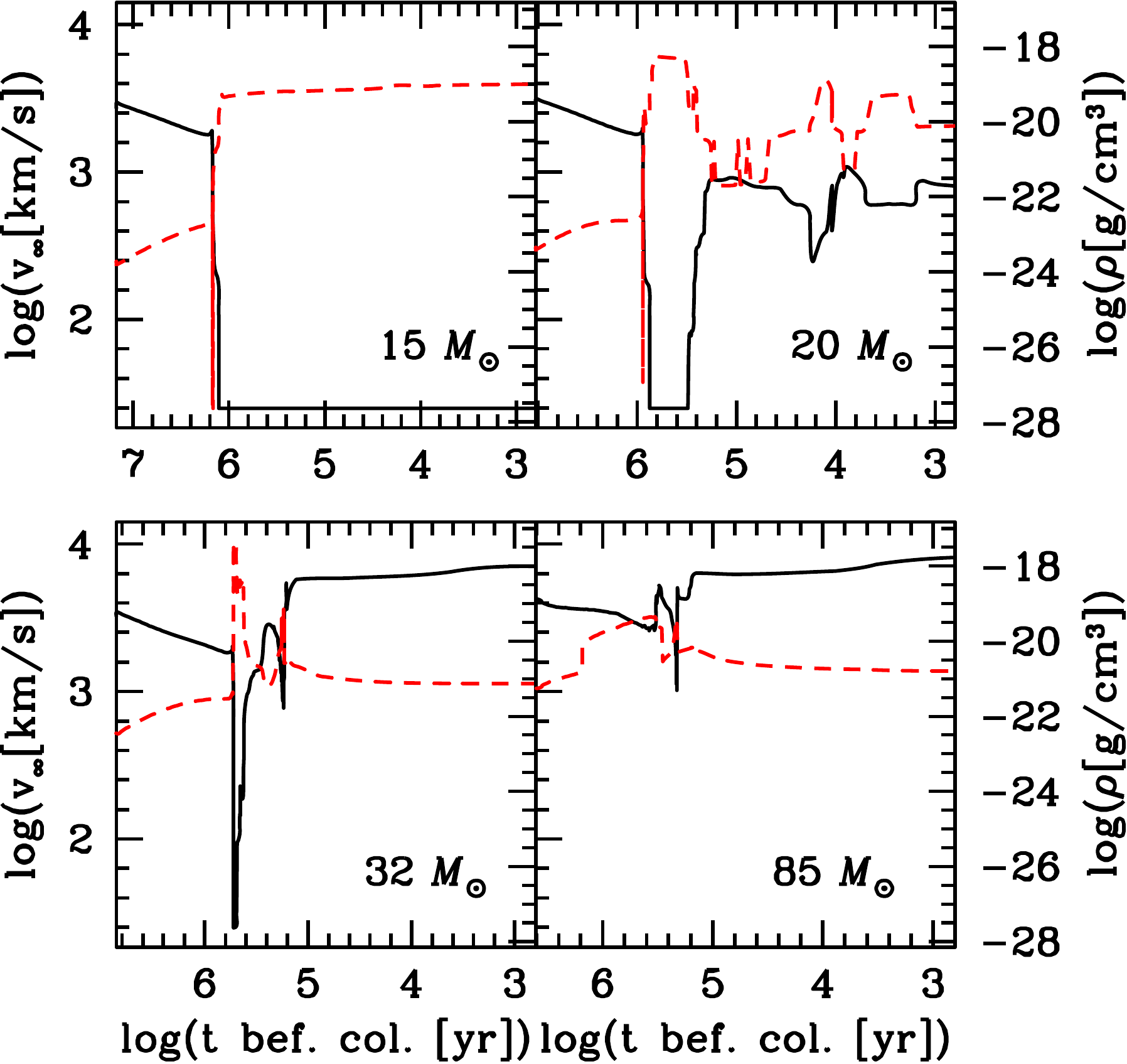}
\end{center}
\caption{Terminal wind velocity (solid black curve) and wind density (dashed red curve) at $r = 10^{16}\, \text{cm}$ from the stellar surface, for the $15$ \textit{(top left)}, $20$ \textit{(top right)}, $32$ \textit{(bottom left)}, and $85\, M_{\sun}$ \textit{(bottom right)} models as a function of the remaining time up to the final collapse.}
\label{FigVitDen}
\end{figure}

Figure~\ref{FigVitDen} shows the time evolution of the terminal wind velocity and density at a distance $r = 10^{16}\, \text{cm}$ from the central star, for the four different mass-loss histories discussed above. For the $15\, M_{\sun}$ star, we see two distinct mass-loss episodes. The first one occurs during the MS with fast radiative winds ($v_\infty\sim 3000\, \text{km}\cdot\text{s}^{-1}$), but with low density $\log(\rho /\text{g}\cdot\text{cm}^{-3})\sim -24$. After the MS, the star becomes a RSG, with very slow winds (set here to $25\,\text{km}\cdot\text{s}^{-1}$), but with a higher density of around $\log(\rho /\text{g}\cdot\text{cm}^{-3})\sim -19$.

The behaviour of the $20\, M_{\sun}$ model is at first similar to that of the $15\, M_{\sun}$. However, after the RSG phase, the star evolves back towards the blue part of the HRD, becoming a WR star. During this last phase, the winds again speed up, while density remains high compared to its value during the MS.

\begin{figure}
\begin{center}
\includegraphics[width=.45\textwidth]{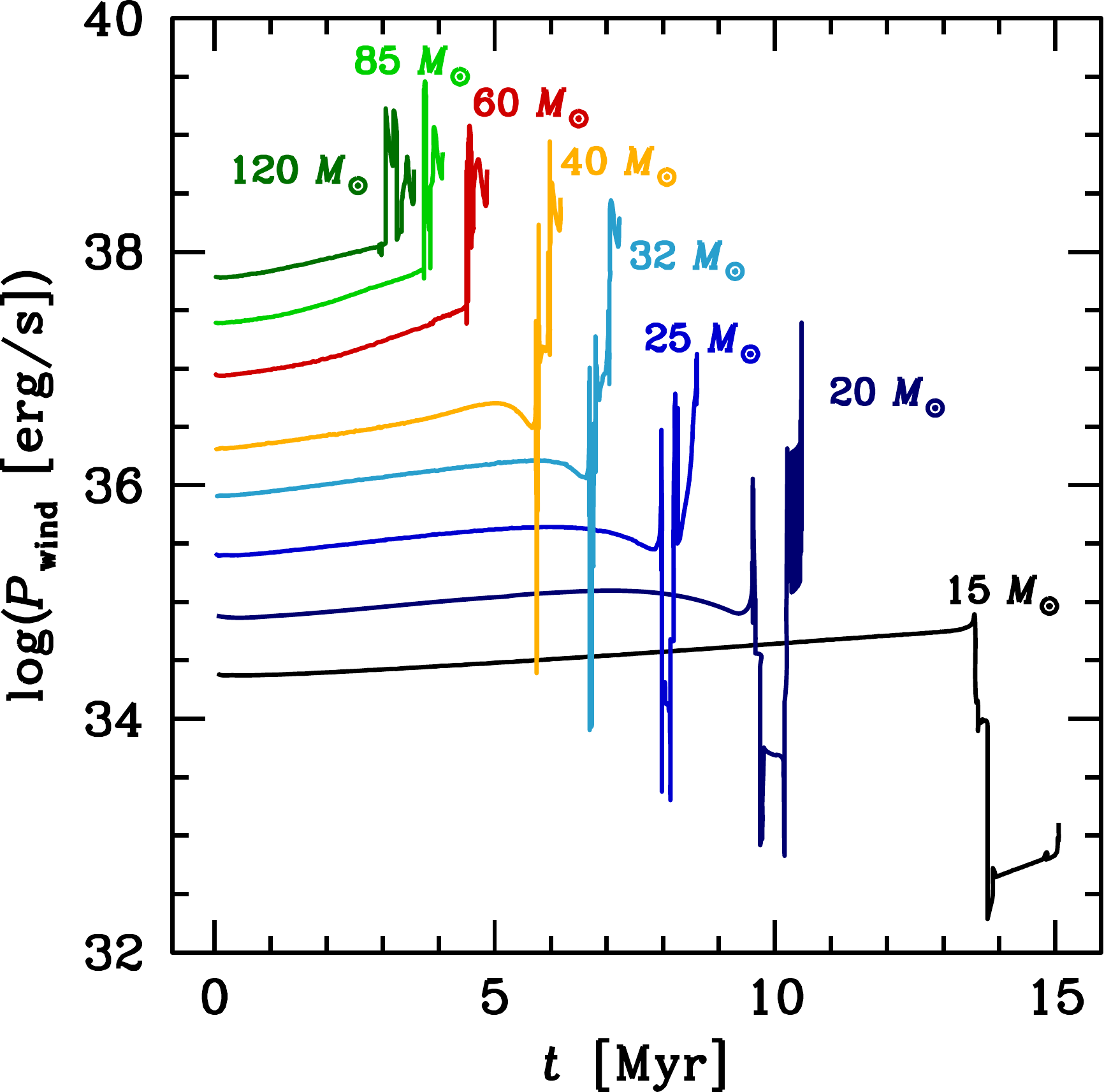}
\end{center}
\caption{Wind power $P_\text{wind} = \frac{1}{2}\dot{M}v_\text{inf}^2$ of the different models (indicated near the curves). The main difference is that in the high-mass models ($M \gtrsim 32\,M_{\sun}$) the wind power slightly increases during the MS, whereas in the lower mass models, the power is close to constant.}
\label{FigWindPower}
\end{figure}

At higher mass, the phase of a slow and dense wind progressively disappears. The star exhibits a very fast and relatively dense wind during the last few $100000$ years before the final collapse. The different and time-dependent mass-loss histories have a strong impact on the power injected in the ISM by the winds. In Fig. \ref{FigWindPower}, the power of the winds of our models is shown as a function of time. During the MS, the wind power is roughly constant for the models between $15$ and $25\,M_\sun$. It is followed by a more or less short period of low-power wind (corresponding to the phase when the star has left the MS, but is still not a WR star), before again entering a strong power wind phase (WR phase, except for the $15\,M_\sun$ model, which avoids this phase). For the more massive models ($\gtrsim 32 \,M_\sun$), the power of the wind increases slightly during the MS (with a possible slight decrease at the very end of the MS), before becoming still stronger during the WR phase.

\section{Details on the hydro-simulations}\label{SecHydroCode}

\subsection{The A-MAZE numerical code and tools}

The simulations were performed with the A-MAZE computational tool box \citep{Walder2000a, Folini2003a}, comprising massively parallel adaptive 3D MHD and radiative transfer codes, together with tools of data analysis and visualisation. This toolbox is validated well in that it has been used for a variety of different astrophysical flow and radiative transfer problems: colliding winds in binaries \citep{Nussbaumer1993a,Walder1998a} and their spectral response \citep{Folini2000a}, wind accretion \citep{Walder2008a}, or supersonic turbulence \citep{Folini2000a,Folini2006a}. For this study, we have added some new tools to the box.

The first new tool is the extension to meshes on general curved manifolds for the case of ideal hydrodynamics (Euler equations). The grid is constructed on the basis of a Cartesian parameter space and a generic, not necessarily differentiable map from this parameter space to the physical space. The advantage of this procedure is that the adaptive mesh algorithm implemented in A-MAZE works also for a mesh on the curved manifold and the same Riemann-solvers as on a Cartesian mesh can be used on the manifold. The implementation closely follows the one described in \citet{Calhoun2008a} and~\citet{Colella1990a}. This new tool of A-MAZE will be presented in detail in a forthcoming paper.

For this study, we used a specific map which results in a mesh identical to a spherical mesh with an equidistant angle discretisation and a logarithmic radial discretisation.  With this mesh, the self-similar evolution phase of the bubble is well recovered, demonstrating the suitability of the chosen approach for the astrophysical problem under consideration (see Fig.~\ref{FigScaling} and the discussion in Sect.~\ref{Sec:Results_Selfsimilarity}).

The second new tool is an elaborate, object-oriented implementation of the description of time-dependent stellar boundary conditions for entire stellar systems. For the present study, it allows us to accommodate in an easy and transparent way the time-dependent boundary conditions of our single star. Although not used in the present study, the implementation can equally handle boundary conditions for entire stellar systems, where each individual star has its own stellar evolution/time-dependent boundary conditions. Also possible here are, again in a transparent way for the user, changes of the outflow characteristics, \textit{e.g.} stellar wind, jet, or supernova explosion. Finally, this new tool can account for proper motions of the stars relative to each other or relative to the computational grid.

From a technical code point of view, a stellar system consists of a given number of stars having different attributes that may be time-dependent: radius, shape, temperature, luminosity, rotation velocity, magnetic fields, etc. Different attributes of (time-dependent) interaction from the stars with their environment can be linked to each star. Among the interaction attributes implemented so far are isotropic and axisymmetric winds, accelerated or launched with the final velocity, jets, nova and supernova explosions. Stellar systems are implemented in three fortran90 modules: one, which defines a stellar system, a star with its attributes and its interactions, by means of fortran90 types; one, which treats the time-dependent boundaries for the different interaction types, and one that links these modules to the general structure of the code. Henceforth, the code is able to automatically change from one star type to another and from one interaction type to another. For instance, if a star explodes as supernova, the star type changes from an ordinary, evolved star to a neutron star or black hole, whereas the interaction type changes from wind to explosion and afterwards to accretion. The object-oriented implementation easily allows to extend the list of stellar or interaction attributes.

The stars may have fixed locations or may move on prescribed orbits or may self-consistently move under the influence of the gravitational forces between them. At the beginning of the simulation, a star track model is loaded. If one wants to follow the evolution of the stellar system after the first explosion of a star, one also loads the explosion characteristics (energy, asphericity, ejected yields, nature of compact remnant, etc.). 

For this first study based on A-MAZE, the system only consists of one star at a fixed position, and the interaction type is always the same: aspherical winds shed with their final velocity. The radius, shape, temperature, luminosity, rotation speed, mass loss, and wind velocity however vary in time. The full power of this new A-MAZE tool can be grasp from Fig.~\ref{FigStellarSystem} in the discussion section.

\subsection{Initial and boundary conditions, numerical settings}

The simulations presented in this paper were performed on a spherical grid, with a logarithmic mesh along the radial direction. The number of cells in the $\theta$-direction is $60$, and in the $r$-direction it varies between 150 and 250, depending on the spatial extension of the physical domain. The radial extension of the domain lies between $[r_\text{min},r_\text{max}]$, where $r_\text{min}$ is chosen to be slightly smaller than $10^{5}$ times the minimal stellar radius during its evolution, and $r_\text{max}$ is chosen large enough to be more than the maximal spatial extension of the bubble. In the $\theta$ direction, the simulations cover the range $[0,\pi]$, with the stellar equator being located at $\theta = \pi/2$. The parameters are summarised in Table~\ref{TabSimSettings}.

\begin{table}
\begin{center}
\caption{Parameters of the numerical grid for each simulation}
\label{TabSimSettings}
\begin{tabular}{c|ccc}
 \rule[0mm]{0mm}{3mm}Simulation $[M_{\sun}]$& $r_\text{min}\,[\text{cm}]$ & $r_\text{max}\,[\text{cm}]$ & resolution\\
\hline
 \rule[0mm]{0mm}{3mm}$15$ & $3\cdot 10^{16}$ & $1.1\cdot 10^{20}$ & 150x60\\
 $20$ & $3\cdot 10^{16}$ & $1.1\cdot 10^{20}$ & 150x60\\
 $25$ & $3\cdot 10^{16}$ & $1.6\cdot 10^{20}$ & 180x60\\
 $32$ & $3\cdot 10^{15}$ & $3.2\cdot 10^{20}$ & 200x60\\
 $40$ & $3\cdot 10^{15}$ & $3.2\cdot 10^{20}$ & 200x60\\
 $60$ & $3\cdot 10^{15}$ & $6.4\cdot 10^{20}$ & 220x60\\
 $85$ & $3\cdot 10^{15}$ & $6.4\cdot 10^{20}$ & 220x60\\
 $120$ & $3\cdot 10^{15}$ & $9.6\cdot 10^{20}$ & 250x60\\
 \end{tabular}
\end{center}
\end{table}

The winds are injected at a distance $r$ equal to $10^5$ times the current stellar radius if this value is less than $0.1\,\text{pc}$, or else at $0.1\,\text{pc}$, with a velocity $v_\infty(\theta)$ and a density $\rho(r,\theta)$ as discussed in Sect.~\ref{SubSecExtraction}. Boundary conditions are reflecting along the symmetry axis (i.e. for $\theta=0$ or $\pi$) and free flow at the outer boundary.

The interstellar medium is generally assumed to have a uniform density of $10^{-24}\,\text{g}\cdot\text{cm}^{-3}$, with some initial random perturbations ($\Delta\rho/\rho = 0.2$). Individual models with higher ISM density are discussed in Sect.~\ref{sec:ism_dens_met}. As the UV-flux of massive stars is very strong, the Str\"omgren radius around our stars is larger than the size of the created bubbles. We thus assume that the ISM is ionised and has a temperature of $10000\,\text{K}$. If during its evolution the effective temperature of the star becomes lower than $10000\,\text{K}$, the ISM and wind can cool down to a temperature of $T_\text{eff}/4$. The chemical composition of the ISM is the same as the initial composition of the stellar wind, i.e. solar: $X=0.72$, $Y=0.266$, and $Z=0.014$, with solar mixture of metals~\citep{Asplund2005a}. We therefore do not consider the structure of the molecular cloud out of which the star is born. Also, all our simulations assume a single star. The effects of binarity or dense clusters will not be considered.

The cooling function used in this work is the one described in \citet{Wiersma2009a}. It allows small modifications in the chemical composition of the medium, which is adapted well to our problem, where the chemical composition of the winds evolves according to the stellar surface. However, the possible range of metallicities allowed by this cooling function is not sufficient to properly treat the most advanced stages of our WR stars. In this case, we use the cooling function with the minimum allowed hydrogen fraction.

\section{Results}\label{SecResults}

\subsection{Circumstellar medium around massive stars for different mass-loss histories}

In this Section, we discuss the broad evolution of the CSM in the various mass-loss history described in Section~\ref{Sec_MassLossHistory}. A more detailed discussion can be found in the next sections.

\begin{figure*}
\begin{center}
\includegraphics[width=\textwidth]{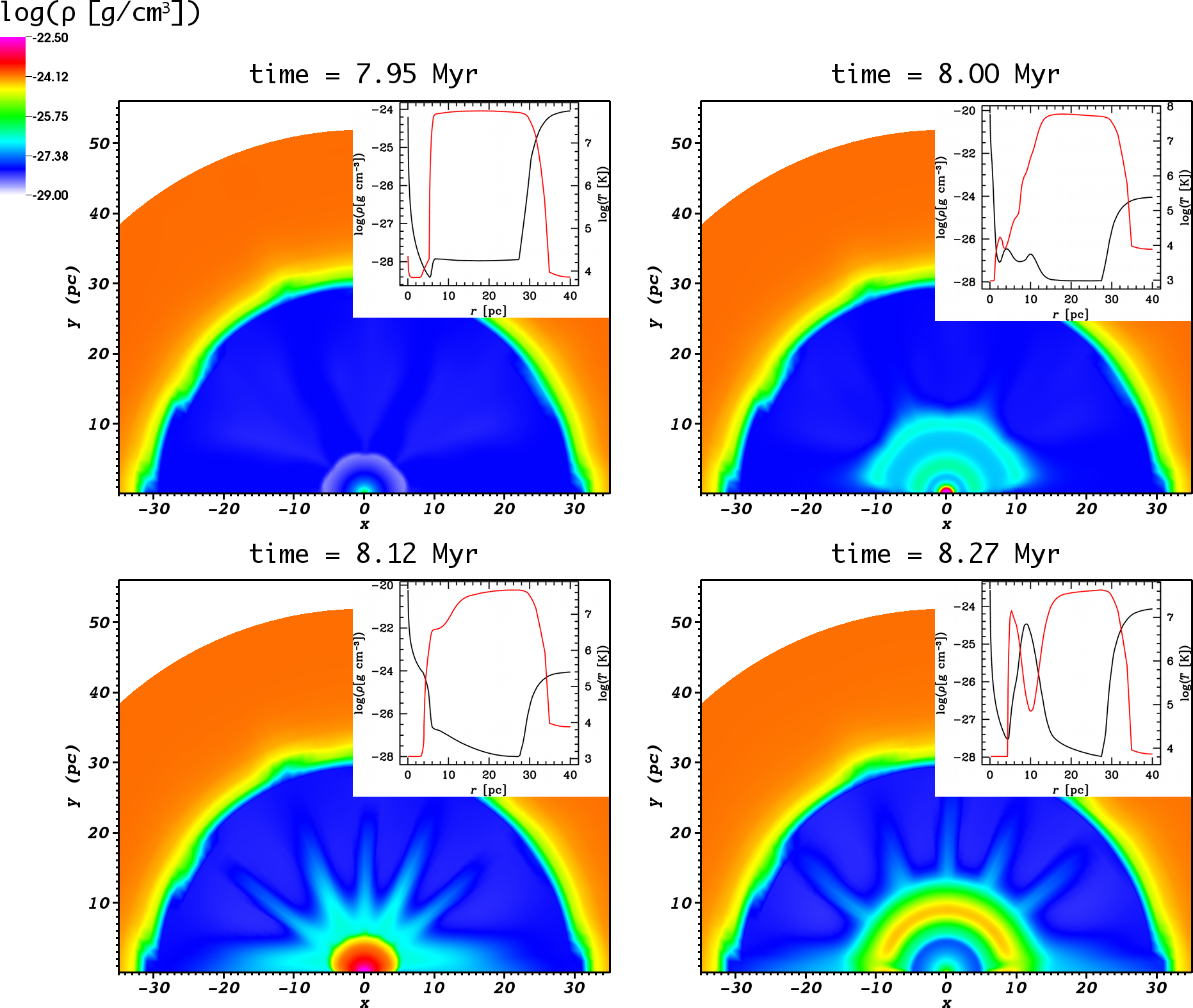}
\end{center}
\caption{Typical aspect of the CSM around a $25\,M_{\sun}$ star at different stages: end of the MS \textit{(top-left panel)}, beginning of the RSG phase \textit{(top-right panel)}, end of the RSG phase \textit{(bottom-left panel)}, and beginning of the WR phase \textit{(bottom-right panel)}. The external circle (at $r\sim 50\,\text{pc}$) is the boundary of the computational domain. The initial density of the ISM is $\log(\rho_\text{ISM}) = -24$. In each panel, we also show in the small window the mean density (black lines) and temperature (red lines) as a function of the radius.}
\label{FigTypicalCSM25}
\end{figure*}

\paragraph{The main sequence phase:} This phase is common and qualitatively similar for all the models computed in the framework of this paper. During this phase, the quick stellar wind (characteristic of massive O-type stars) progressively pushes the ISM away. The size of the bubble increases progressively during the MS. Unlike other similar studies \citep{Garcia-Segura1996a,Garcia-Segura1996b}, where a strong over-density (compared to the density of the ISM) is observed at the outer edge of the bubble, we do not obtain such a structure in our simulations. We come back to this point in more detail in Sect.~\ref{sec:nebulae} and only stress here that our result is coherent with the recent study by \citet{Toala2011a}, who accounted for the effects of ionisation and found an over-density only further away from the star at the location of the ionisation front. The typical aspect of the CSM at the end of the MS is shown in Fig.~\ref{FigTypicalCSM25} \textit{(top-left panel)}. Near the star, the density of the stellar wind decreases with increasing radius, up to the wind-termination shock, where the density is minimal. In the bubble, the density is then roughly constant, up to the edge with the ISM.

\begin{figure*}
\begin{center}
\includegraphics[width=\textwidth]{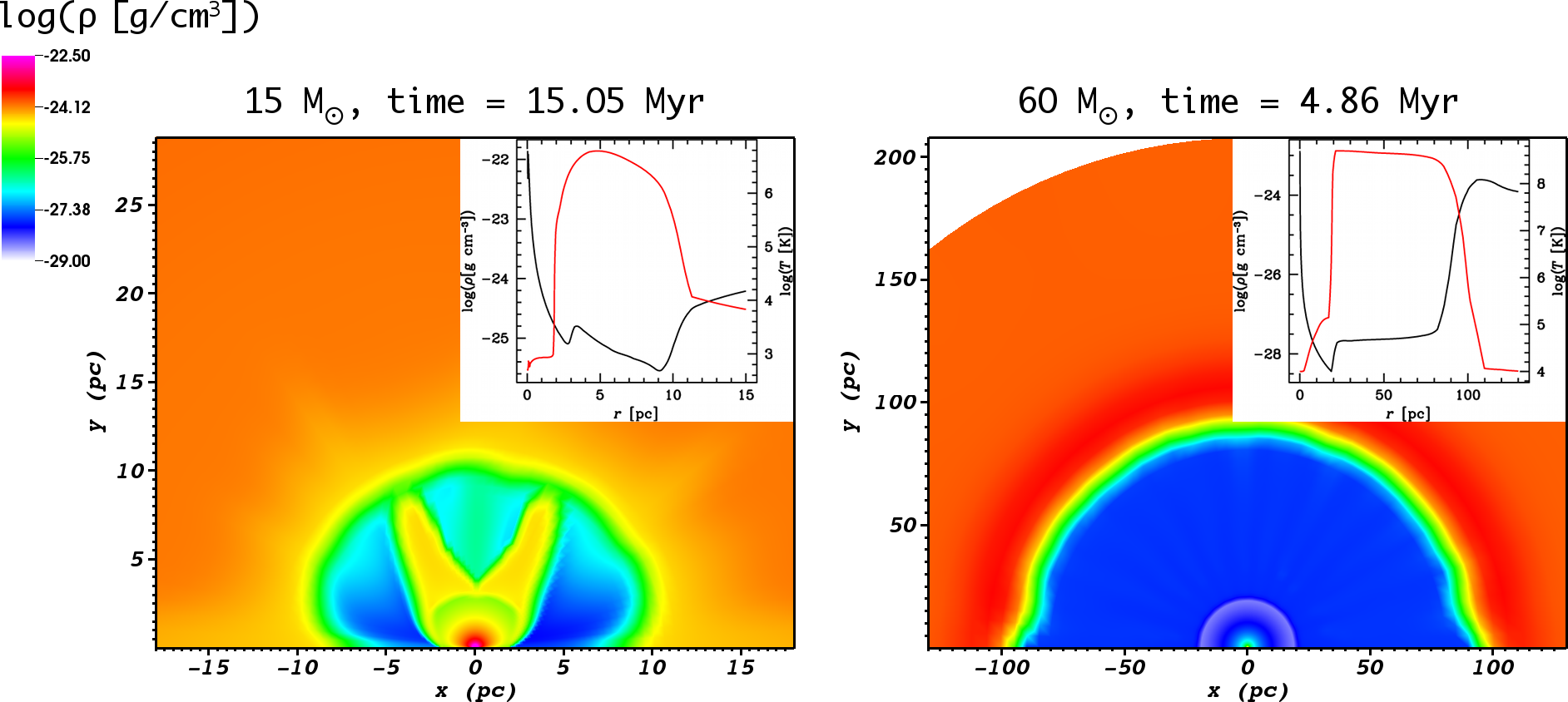}
\end{center}
\caption{Typical pre-SN aspect of the CSM around a $15\,M_{\sun}$ star \textit{(left panel)} ending its stellar life as a RSG (and exploding as a type IIP SN), and around a $60\,M_{\sun}$ star \textit{(right panel)}, ending as a WR star (exploding as a type Ibc SN). In each panel, we also show the mean density and temperature in the small window as a function of the radius.}
\label{FigTypicalCSMEnd}
\end{figure*}

\paragraph{The O-star -- RSG scenario:} This evolutionary path is followed by models with an initial mass $M_\text{ini} \leq 15\, M_{\sun}$. After the MS phase, these stars enter the RSG phase, characterised by a slow and dense wind. This wind creates a dense shell around the star, which progressively extends through the bubble left after the MS (see Fig.~\ref{FigTypicalCSM25}, \textit{top-right} and \textit{bottom-left panel}). In the transition region from the slow RSG wind to the fast MS wind, the velocity difference of the two winds results in a rarefaction wave, possibly accompanied by a shock. Instabilities are generated from numerical seeds as these waves hit the previously existing wind termination shock~\citep[e.g.][]{Walder1998b}. The associated slight density enhancements are further amplified and, to some degree, preserved by cooling effects. Rayleigh-Taylor instabilities of the RSG shell are also likely to be present \citep{Dwarkadas2007a}. Details of the instabilities are washed out by the coarse resolution of the simulation, so what remains are the over-dense rays apparent in Fig.~\ref{FigTypicalCSM25}. Also at the onset of the slow wind, the pressure inside the bubble decreases slightly, causing the bubble to shrink. The high density eases the cooling, and the dense matter shell is relatively cold compared to the rest of the bubble, with temperatures of the order of $10^{4}\,\text{K}$. At the end of its life, the star is enshrouded in an asymmetrical and dense region, surrounded by the remnant of the bubble created during the MS (Fig.~\ref{FigTypicalCSMEnd} \textit{left panel}). The high-density structure seen in the interior of the bubble is the result of two large roll-ups that ultimately gain over other, higher order modes that are initially present. A full 3D simulation would certainly show a different structure.

\paragraph{The O-star -- RSG -- WN scenario:} In the framework of the new grids of stellar models from the Geneva group \citep{Ekstrom2012a}, the models with an initial mass $M_\text{ini}\sim 20-25\, M_{\sun}$ again cross the HRD after becoming a RSG as described above, and end their stellar lives as WN stars. In that case, after a short episode of slow wind, the stellar wind becomes much more rapid. As shown in Fig.~\ref{FigTypicalCSM25} \textit{(bottom-right panel)}, the WR wind progressively pushes the dense shell produced during the RSG wind episode away. This shell progressively dilutes in the previously built bubble, increasing its density and softening the edge of the interface between the bubble and the ISM. Near the central star, a structure similar to the one existing during the MS appears, with a shock braking the wind of the WR star. 

\paragraph{The O-star -- YSG -- WN -- WC scenario:} This kind of evolution is typical of stars with initial mass in the range $\sim 32\, M_{\sun}$ to $\sim 40\, M_{\sun}$. In that case, at the end of the MS, the star's effective temperature decreases briefly down typically to $\log(T_\text{eff}) \sim 3.8$ (and thus completely avoiding the RSG branch). During this short phase (during which the star is a yellow supergiant, YSG), the wind density increases, and the wind velocity decreases, but not as much as in the RSG case. However, the effect is relatively similar, with the ejection of a dense and cold shell around the star, which is progressively expelled away by the upcoming very fast wind of the WR star. We do not observe instabilities leading to the formation of a thin and dense shell, as usually reported by other authors \citep[for example,][]{Garcia-Segura1996b}. As mentioned in \citet{vanMarle2012a}, this is certainly because our spatial resolution is not high enough to capture this behaviour. The final result is that the mean density of the bubble is slightly increased by the dilution of the slow wind and that the edge of the cavity is not as sharp as during the MS.

\begin{figure*}
\begin{center}
\includegraphics[width=.48\textwidth]{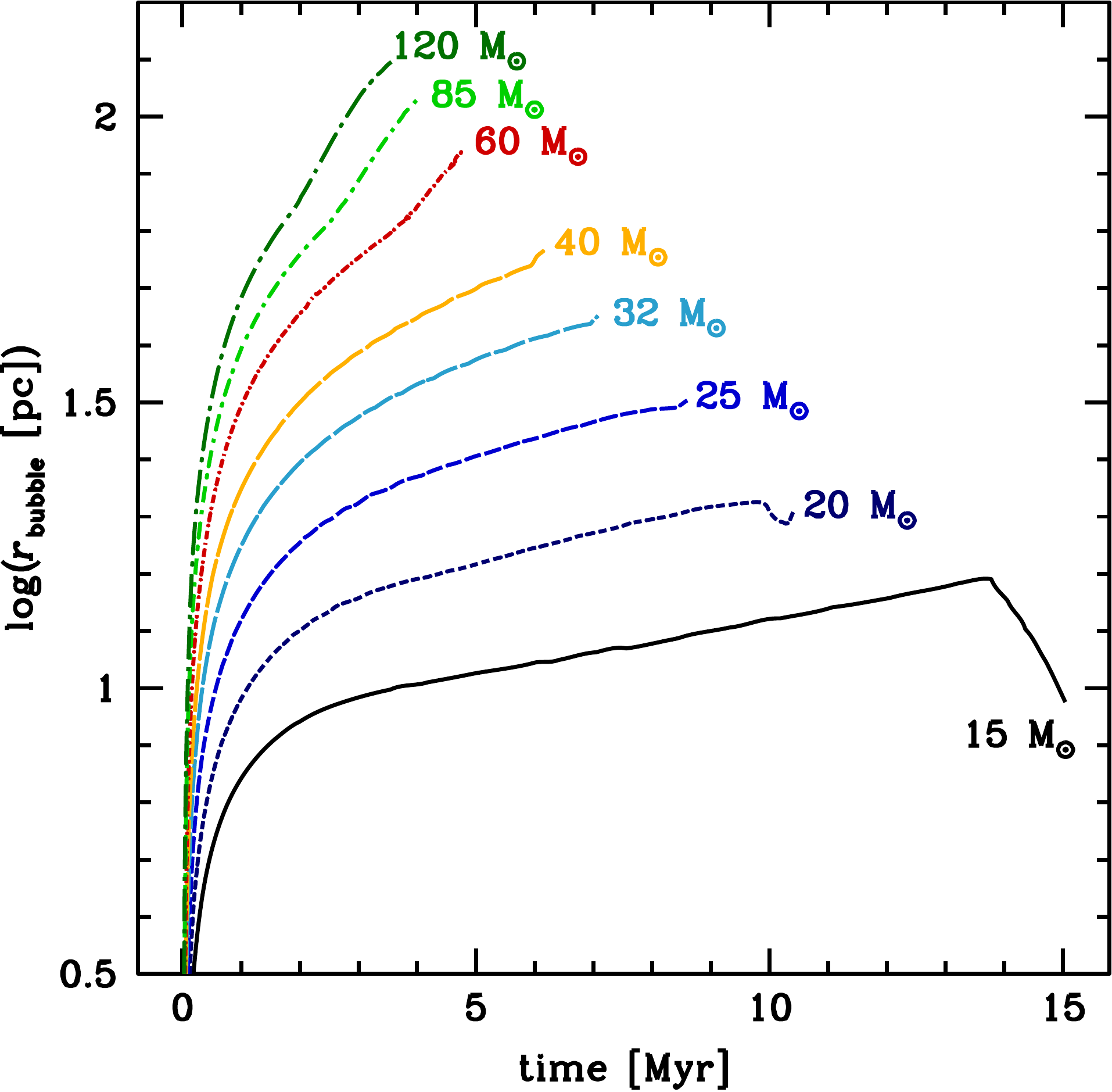}\hfill\includegraphics[width=.48\textwidth]{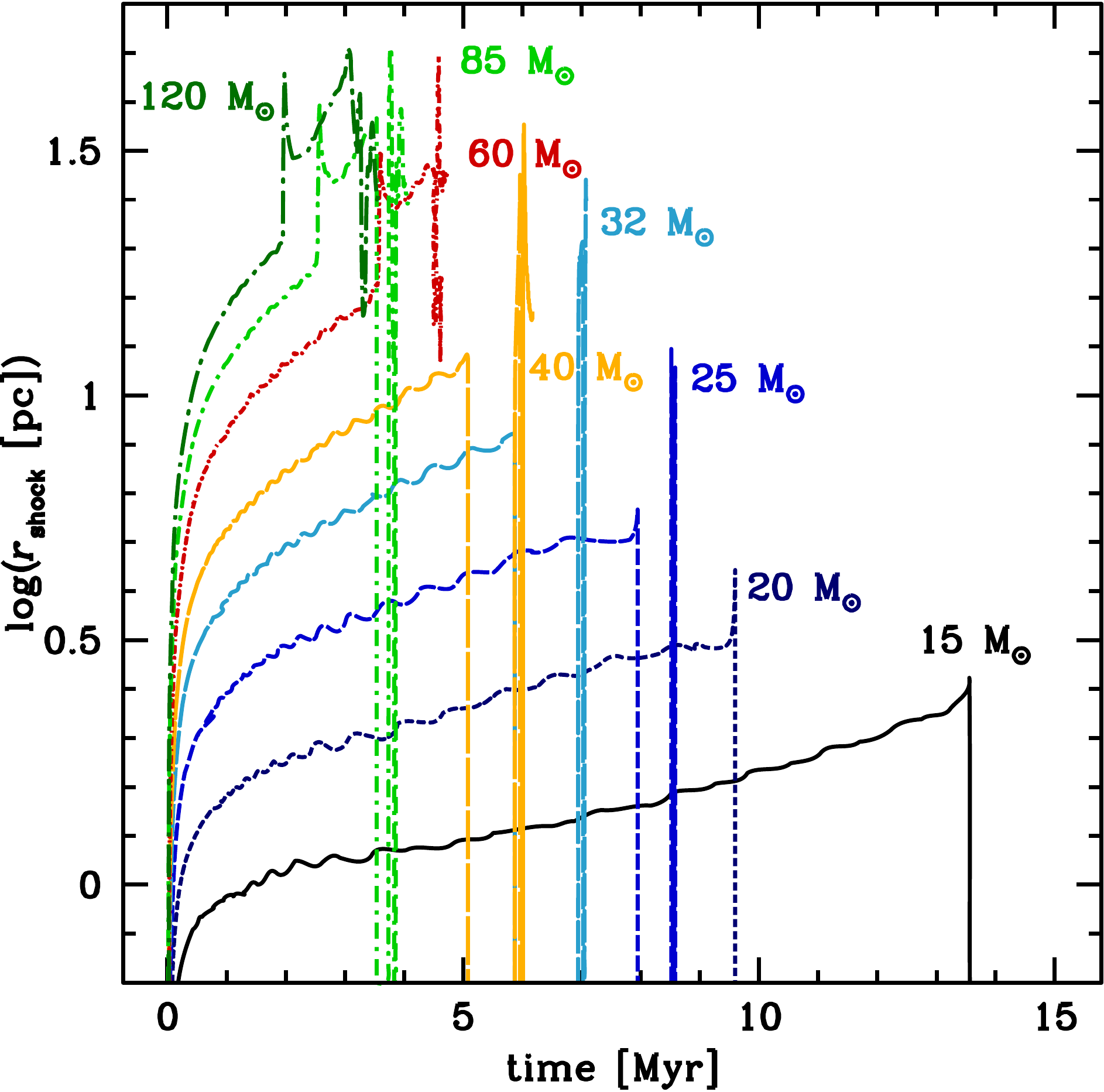}
\end{center}
\caption{Radius of the bubble \textit{(left panel)} and position of the wind termination shock \textit{(right panel)} for our models.}
\label{FigRadii}
\end{figure*}

\paragraph{The O-star -- WN -- WC scenario:}  For the most massive stars, the WR phase begins before the end of the MS. In that case, the star remains at very high $T_\text{eff}$ during its entire evolution. Particularly, there is no ejection of a dense and slow wind before the onset of the WR wind. The evolution after the MS is thus in the continuity of the first time evolution, except that the bubble growth velocity increases slightly, because the WR wind deposits more momentum. The typical aspect of the bubble at the end of the star's life is shown in Fig.~\ref{FigTypicalCSMEnd} \textit{(right panel)}.

\subsection{Size and chemical composition of the bubble}

The left-hand panel of Fig.~\ref{FigRadii} shows the size of the bubble, defined as the region where the medium consists of at least $90\%$ stellar material and the position of the wind termination shock (\textit{right panel}). Thanks to the increasing mass-loss rate and wind velocity with increasing stellar mass, the size of the bubble is larger for higher initial mass. The final size of the bubble spans from $\sim10\,\text{pc}$ for the bubble around the $15\,M_{\sun}$ model to more than $100\,\text{pc}$ around the $120\,M_{\sun}$ model. Near the end of the tracks of the $15-25\,M_{\sun}$ models, a decrease in the size of the bubble is apparent, which occurs during the RSG phase. During this phase, the pressure in the bubble decreases, causing it to shrink. For the models ending as a WR star, the growth of the bubble increases at the onset of the fast and dense WR wind.

The position of the wind termination shock follows the same trend. Because it depends strongly on the wind parameters (mass-loss rates and velocity), which can change slightly during the stellar evolution, it is not perfectly smooth during the MS. It disappears during the RSG phase for models having such a stage. During the WR phase, owing to the complexity of the CSM near the star, it is sometimes difficult to attribute a position to this shock. This explains the somewhat chaotic curves during this phase. For the most massive models, the position of this shock is mostly farther than $10\,\text{pc}$ from the star.

\begin{figure}
\begin{center}
\includegraphics[width=.48\textwidth]{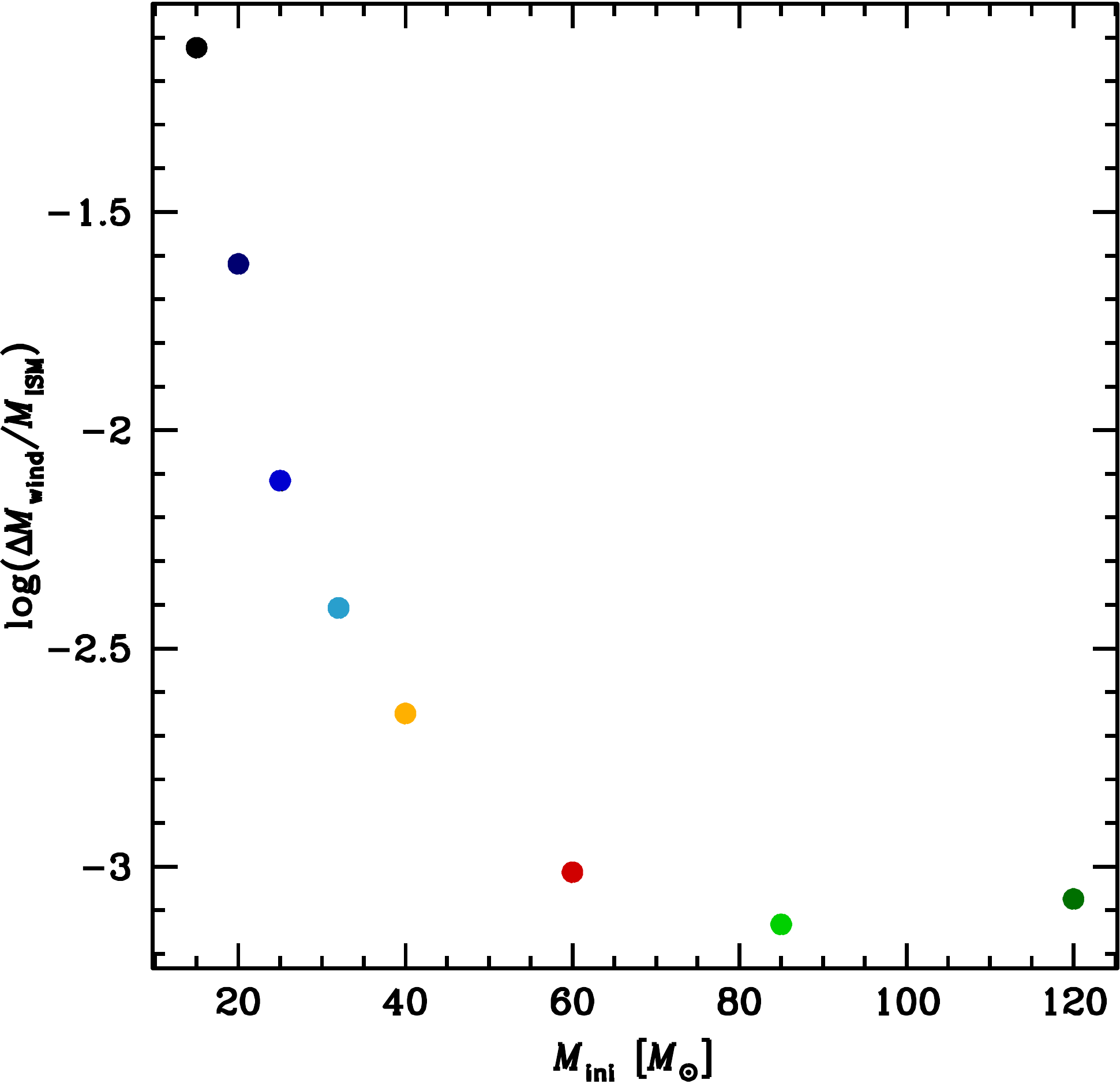}
\end{center}
\caption{Ratio of the mass shed by the stellar wind to the mass of the ISM pushed away right before the SN explosion. This is representative of the density contrast between the mean density in the bubble and the ISM density (see text). Colours as in Fig.~\ref{FigRadii} denote the different models.}
\label{FigMassRatio}
\end{figure}

An interesting quantity is the ratio between the amount of mass lost by the star during its entire life, $\Delta M_\text{wind}$, and the amount of interstellar medium mass pushed away by the stellar wind during this time, $M_\text{ISM}$. This ratio, shown in Fig.~\ref{FigMassRatio} for all models, is representative of the density contrast between the mean density in the bubble and the ISM density since most of the matter present in the bubble originates in the stellar winds. As can be seen, the mean density in the bubble is the lowest for the most massive models. The mean density in the bubble around the $15\, M_{\sun}$ model is roughly one tenth of the initial ISM density. For the  $120\, M_{\sun}$ model, this number drops to less than $10^{-3}$. Although $\Delta M_\text{wind}$ is highest for the most massive models, the increase in bubble size with stellar mass compensates for and even dominates it, explaining this contrast.

Owing to the action of mass loss and mixing, the chemical composition of the stellar wind changes progressively as a function of time \citep[see][their Fig.~6]{Georgy2012b}. This produces a progressively inhomogeneous modification of the chemical composition of the CSM in the bubble. Observationally, several measurements show that most of the WR bubbles are enriched in He and N \citep{Esteban1992a}, typically corresponding to matter processed by the CNO cycle. More recently, \citet{Fernandez-Martin2012a} have measured various abundances at different locations in the WR nebula NGC 6888 (the central star is a WN6 star) and found different compositions in different regions. They particularly found that the WR nebula can be divided into three zones, with different chemical compositions: an inner shell showing a strong nitrogen enrichment, an O/H ratio slightly deficient (compared to the solar value), and a strong overabundance of He. This shell is surrounded by an outer shell, where He/H abundance is higher than the solar value, but where both N/H and O/H do not seem to be enriched. Finally, there is a surrounding skin, with composition similar to the one expected for the ISM at this galactic radius.

\begin{figure}
\begin{center}
\includegraphics[width=.48\textwidth]{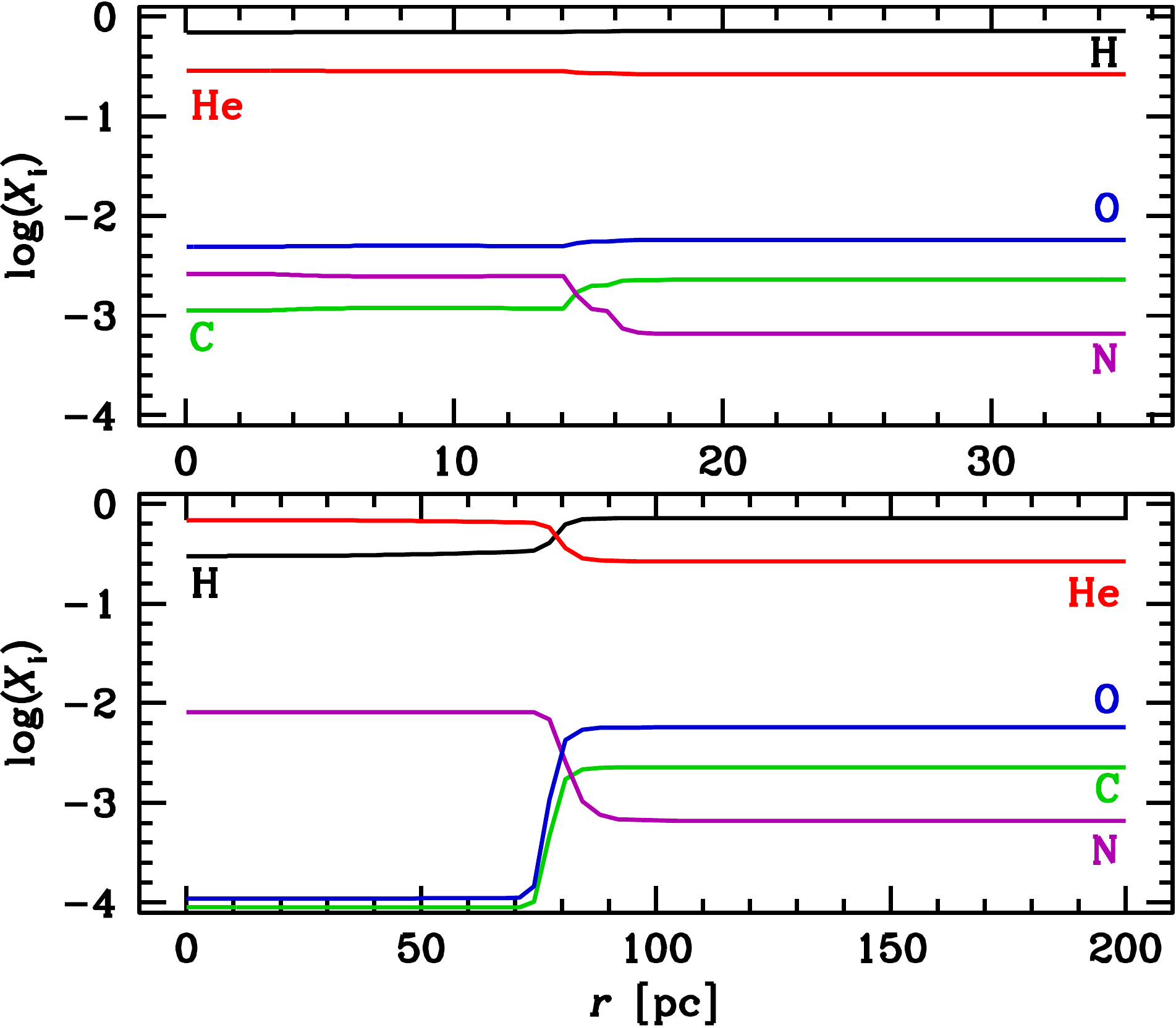}
\end{center}
\caption{Chemical abundances (mass fractions) as a function of the radius around a $15\, M_{\sun}$ star at the end of the MS \textit{(top panel)} and a $120\, M_{\sun}$ star \textit{(bottom panel)} immediately before becoming a WR star. The bubble radius is $9.5\,\text{pc}$ for the $15\, M_{\sun}$ model and $83\,\text{pc}$ for the $120\, M_{\sun}$.}
\label{FigAbundEndMS}
\end{figure}

In our simulations, the distribution of chemical species around the star during its life can vary a lot and lead to very inhomogeneous structures. In Fig.~\ref{FigAbundEndMS}, we show the situation at the end of the MS (for a $15\, M_{\sun}$ star, \textit{top panel}) and immediately before entering the WR phase (for a $120\, M_{\sun}$ star, \textit{bottom panel}). The compositions shown are averages over constant radii (x-axis). Both cases are quite similar, the chemical composition in the bubble being almost constant (except near the edge of the bubble) and showing the products of CNO-cycle burning at low temperature (increase in He and N, decrease in H and C, O almost constant) for the $15\, M_{\sun}$ model, and at high temperature for the $120\, M_{\sun}$ (decrease in H, C, and O, increase in He and N).

\begin{figure}
\begin{center}
\includegraphics[width=.48\textwidth]{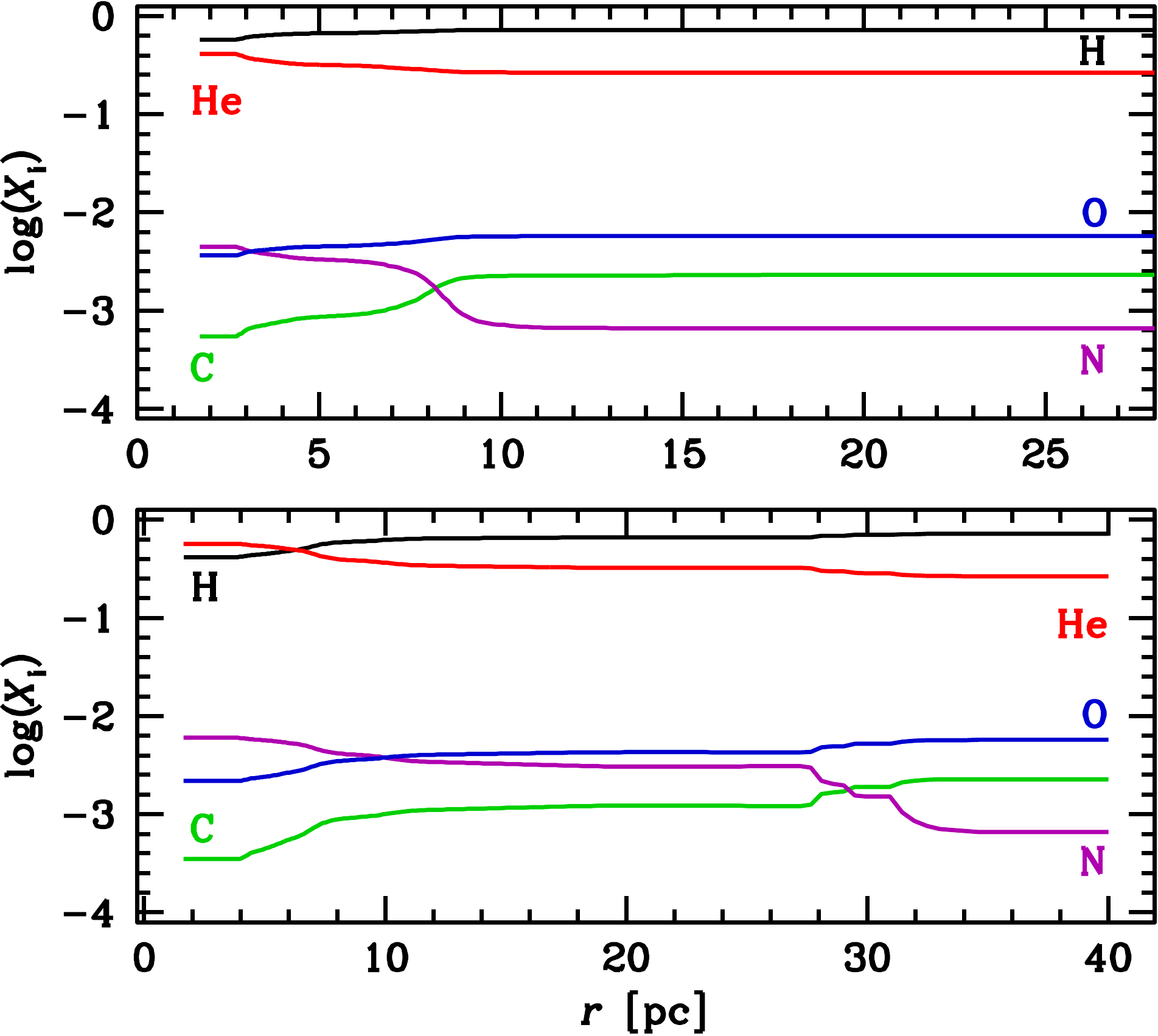}
\end{center}
\caption{Chemical abundances (mass fractions) as a function of the radius around a $15\, M_{\sun}$ star \textit{(top panel)} and a $25\, M_{\sun}$ star \textit{(bottom panel)}  at the end of the RSG phase. The bubble radius is about 10 pc for the $15\, M_{\sun}$ model and 32 pc for the $25\, M_{\sun}$ model.}
\label{FigAbundEndRSG}
\end{figure}

For stars having a RSG phase after the MS, the situation evolves slowly during this phase with slow and dense winds. The development of a large external convective zone below the stellar surface means that more CNO-cycle elements are rapidly brought to the surface and carried away by the winds. This is illustrated in Fig.~\ref{FigAbundEndRSG}, where we see that the chemical composition in the external zone of the bubble remains similar to the one at the end of the MS, while near the star, where the shell of dense matter produced by the RSG winds is located, we observe a medium enriched in He and N and depleted in H, C, and O compared to the external zone.

\begin{figure}
\begin{center}
\includegraphics[width=.48\textwidth]{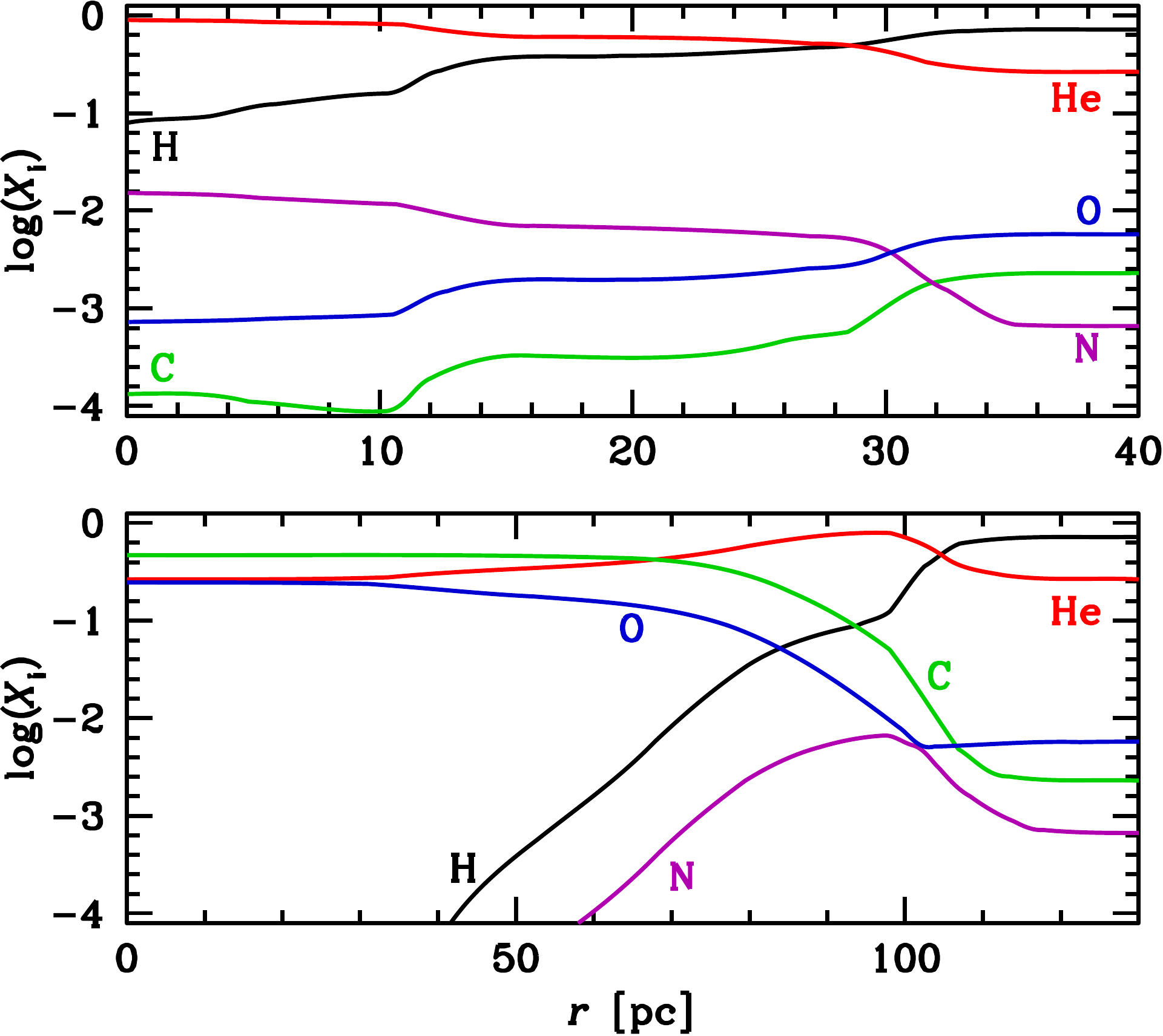}
\end{center}
\caption{Chemical abundances (mass fractions) as a function of the radius around a $25\, M_{\sun}$ star \textit{(top panel)} and a $85\, M_{\sun}$ star \textit{(bottom panel)}  at the end of the WR phase. The bubble radius is about 32 pc for the $25\, M_{\sun}$ model and 100 pc for the $85\, M_{\sun}$ model.}
\label{FigAbundEndWR}
\end{figure}

\begin{figure*}
\begin{center}
\includegraphics[width=.49\textwidth]{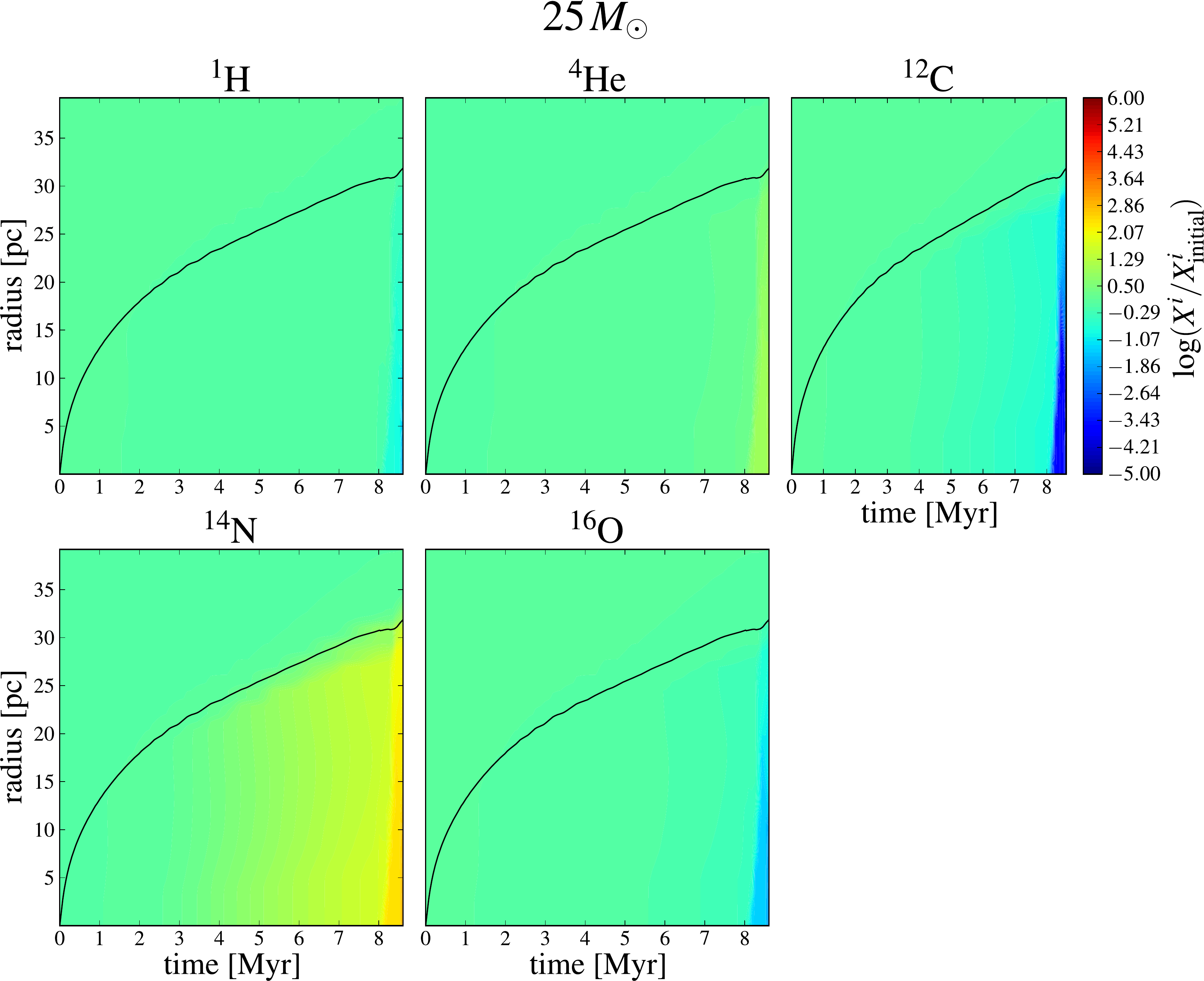}\hfill\includegraphics[width=.49\textwidth]{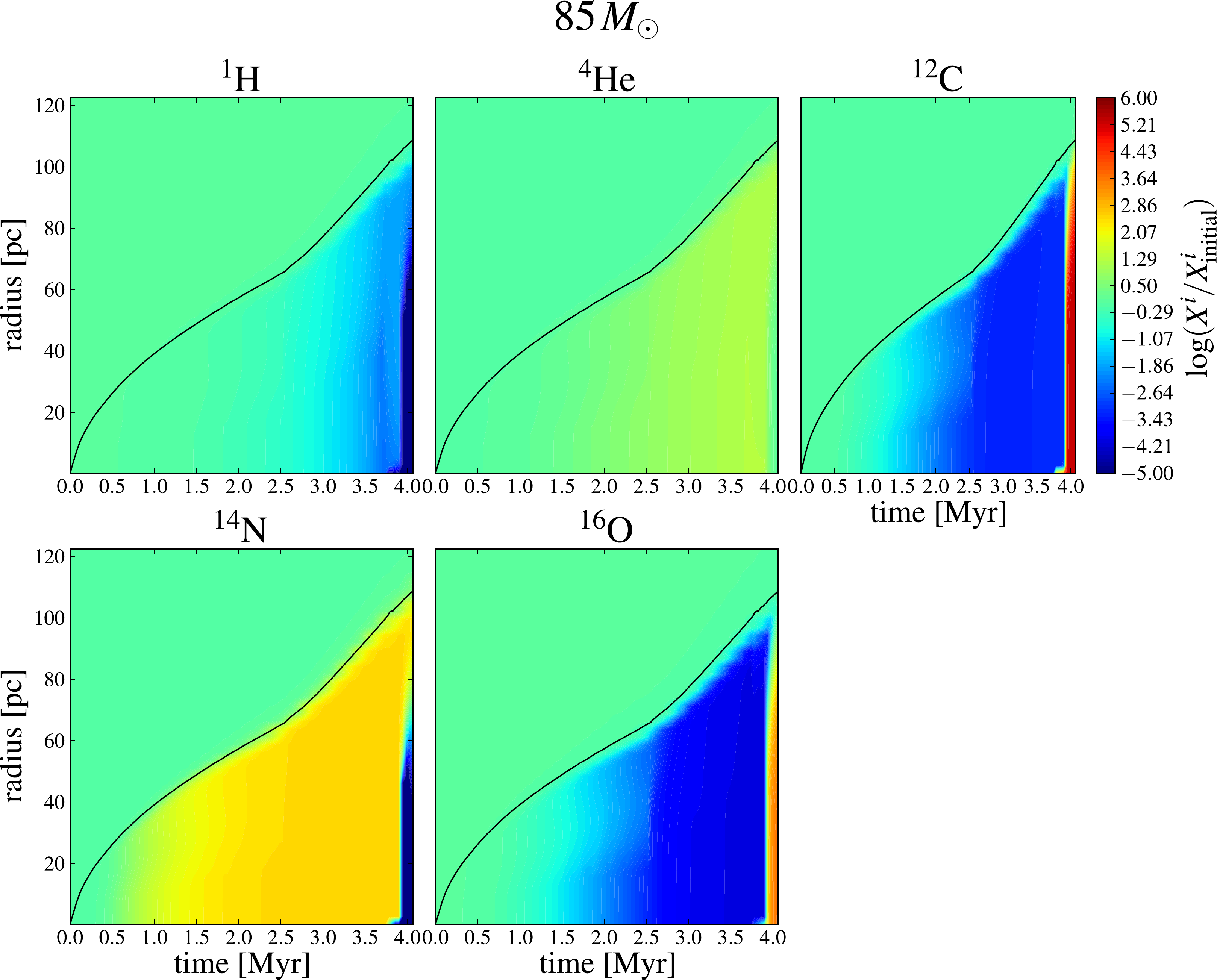}
\end{center}
\caption{Evolution of the chemical composition of the bubble around the $25\,M_\sun$ (\textit{left panel}) and $85\,M_\sun$ (\textit{right panel}). Each followed chemical element is shown in its own panel, with respect to the initial values. The red regions are regions where the element mass fraction is 1 million times higher than the initial one, and the blue regions where the element mass fraction is 100'000 times lower than initially. The black line shows the edge of the bubble.}
\label{FigChemComplet}
\end{figure*}

Above $20\, M_{\sun}$, the stars have a WR phase at the end of their lives. Figure~\ref{FigAbundEndWR} shows the abundance profile in the bubble for the $25\, M_{\sun}$ and $85\, M_{\sun}$ models. The first one evolves into a WR star after a RSG stage, while the second already becomes a WR star during the MS, avoiding the RSG stage altogether. In the case of the $25\, M_{\sun}$ star, we see that the abundance profile at the end of the WR phase has changed since the end of the RSG stage (see the bottom panel of Fig.~\ref{FigAbundEndRSG}). The stellar winds have progressively filled the bubble with material strongly modified by the CNO-cycle burning: a strong H depletion, with He becoming the dominant element in a large fraction of the bubble. In the meantime, C and O abundances decrease, and are replaced by N. A complete view of the evolution of chemical species in the bubble of the $25\,M_\sun$ model is shown on the left-hand panel of Fig.~\ref{FigChemComplet}.

The case of the $85\, M_{\sun}$ is even more extreme. The very strong mass loss enriches the wind not only in the burning products of hydrogen, but also in the products of He-burning (mostly C and O) at the end of the evolution. This creates two distinct regions in the bubble. In the external region (between $80$ and $110\, \text{pc}$), we find the H-burning products. Particularly, the He and N abundances are maximal; however, this area is also polluted in C and O from the internal region, increasing the abundance of these elements. In the inner region, C becomes the dominant species. Oxygen is also strongly enhanced, while He is depleted. Finally, H and N are completely absent in this region. The time evolution as a function of the radius for each element illustrating these various effects is shown in the right-hand panel of Fig.~\ref{FigChemComplet}. Looking at both panels of the same figure, we also see that the stellar winds are efficient to quickly modify the chemical composition inside the bubble, the abundances variation following the chemical composition of the stellar surface quite closely.

The case of NGC 6888 and its central star WR 136 \citep{Johnson1965a} qualitatively fits our results. According to the stellar parameters given in \citet{Hamann2006a} and to the HRD tracks by \citet{Ekstrom2012a}, the central WR star (a WN6 star) corresponds to a star with initial mass around $32\, M_{\sun}$. During the WN phase of this model, the abundance profile is similar to the $25\, M_{\sun}$ one (top panel of Fig.~\ref{FigAbundEndWR}), with a central region strongly modified by the products of the H burning (He and N increased, H, C, and O depleted). Near the edge of the bubble, we find a transition region towards the composition of the ISM.

\subsection{ISM density and metallicity effect}
\label{sec:ism_dens_met}

The final radius of the bubble, hence the size of the region where the stellar winds substantially modify the chemical composition of the medium, mainly relies on the assumed density of the ISM and on the efficiency of the stellar winds at injecting momentum into the circumstellar medium. This last point itself depends on the metallicity of the central star, since the radiative mass-loss rate depends on the metal content of the stellar atmosphere \citep[$\dot{M}\sim Z^{0.85}$,][]{Vink2001a}\footnote{\footnotesize{Changing the metallicity of the star not only changes the total amount of mass lost, but also affects substantially the evolutionary tracks in the HRD.}}.

To quantify these effects, we computed six additional models: four $25$ and $60\, M_{\sun}$ models at $Z_{\sun}$, with $\log(\rho_\text{ISM} [\text{g}\cdot \text{cm}^{-3}]) = -23$ (10 times our standard value) and $\log(\rho_\text{ISM} [\text{g}\cdot \text{cm}^{-3}]) = -22$ (100 times the standard value), and two models at the metallicity of the Small Magellanic Cloud (SMC, $Z=0.002$), one of $25\, M_{\sun}$ and one of $60\, M_{\sun}$ \citep[][in press]{Georgy2013b}. The final bubble radii of these models are summarised in Table~\ref{TabRhoZ}. Increasing the ISM density decreases the size of the bubble. The multiplying factor is dependent on the initial mass ($3.2-3.6$ for the $25\,M_\sun$ model, $1.9-2.2$ for the $60\,M_\sun$). A more complete discussion can be found in Sect.~\ref{Sec:Results_Selfsimilarity}. Decreasing the metallicity by a factor of 7 (between $Z_{\sun} = 0.014$ and $Z_\text{SMC} = 0.002$) decreases the size of the bubble by a factor of $\sim 1.7$.

\begin{table}
\begin{center}
\caption{End MS and final bubble radii for models with modified $\log(\rho_\text{ISM})$ and $Z$.}
\label{TabRhoZ}
\begin{tabular}{cc|ccccc}
\rule[0mm]{0mm}{3mm}$M_\text{ini}$ & &  $\log(\rho) =$ & $-24$ & $-23$ & $-22$ & $-24$\\
\rule[0mm]{0mm}{3mm}$M_{\sun}$& & $Z=$ & $Z_{\sun}$ & $Z_{\sun}$ & $Z_{\sun}$ & $Z_\text{SMC}$\\
\hline
\rule[0mm]{0mm}{3mm}$25$ & end MS & & $30.73$ & $9.66$ & $2.92$ & $18.40$\\
$25$ & final & & $31.84$ & $9.06$ & $2.79$ & $18.81$\\
$60$ & end MS & & $79.47$ & $40.77$ & $17.07$ & $48.68$\\
$60$ & final & & $88.76$ & $45.54$ & $20.62$ & $50.21$\\
\end{tabular}
\tablefoot{All radii are given in pc.}
\end{center}
\end{table}

These results show the extreme dependence of this kind of simulation on the various initial and boundary conditions. Peculiar results obtained for one stellar model of a given mass and metallicity are hardly applicable to other situations. This demonstrates the value of grids of models as presented here. 

Another interesting question in this context is what the evolution of the bubble in a very inhomogeneous medium would be, which is a typical situation for star-forming regions where most of the massive stars are found. The simulations presented in this work are valid for a single massive star, and do not take possible binarity or the presence of one or more other massive stars in the surrounding few 10 pc into account, as would be the case in a cluster, for example. In the latter case, the situation would be much more complex, with interactions between the stellar winds. Some exploratory work into this direction is shown in Fig.~\ref{FigStellarSystem}, where we see complex turbulent interaction regions of the individual stellar winds. Mixing of chemical species is probably more efficient in such interaction regions, as is the outward transport of matter and momentum: the interaction regions in Fig.~\ref{FigStellarSystem} extend farther outward than do the (partial) bubbles of the individual stars \citep[see also][]{vanMarle2012b}. The possibility that the star is moving with respect to the ISM also changes the behaviour of the CSM \citep[see for example the recent work by][]{Decin2012a,Mackey2012a}.

\subsection{Nebulae around O- and WR-stars}
\label{sec:nebulae}

\begin{figure}
\begin{center}
\includegraphics[width=.48\textwidth]{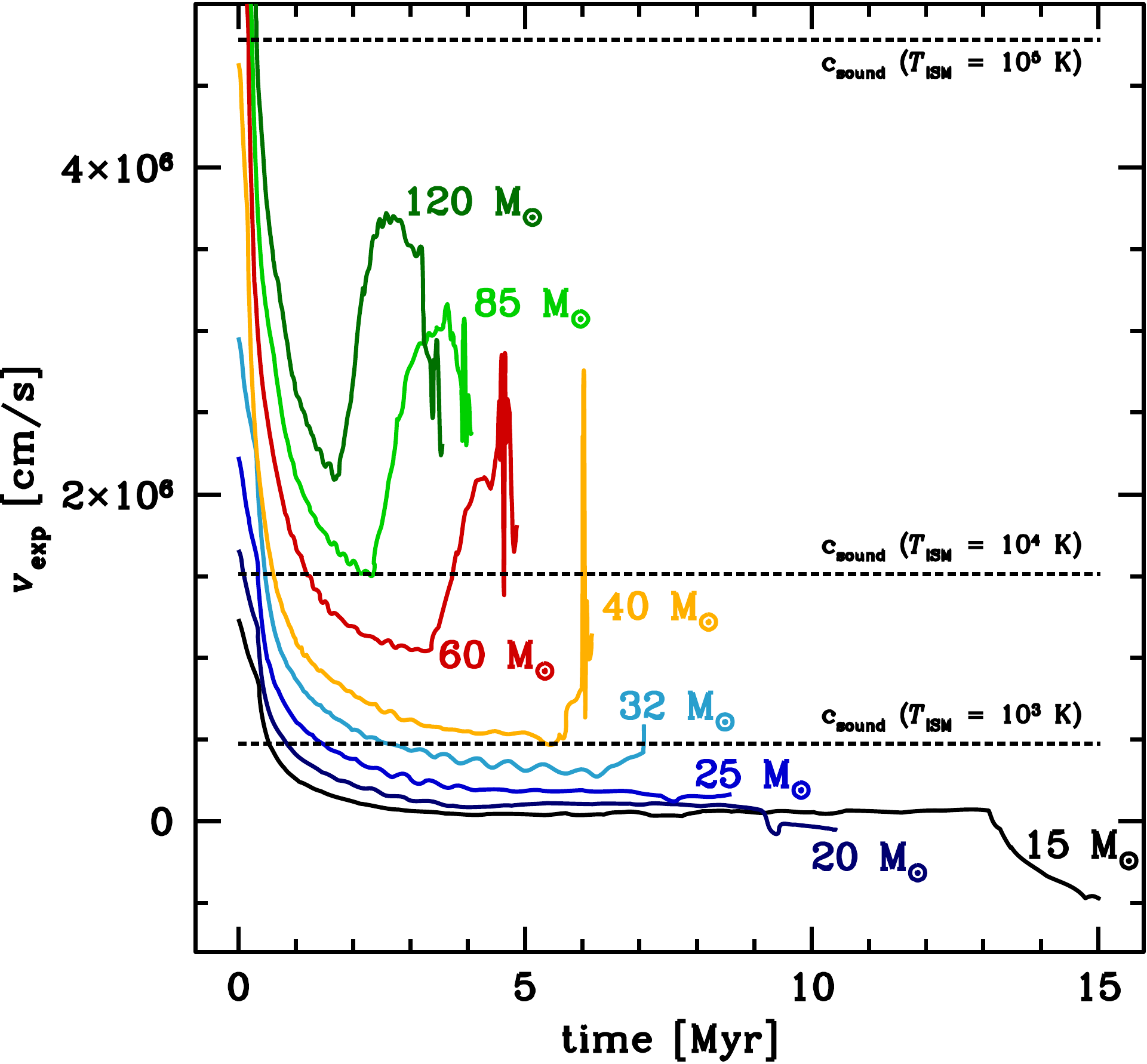}
\end{center}
\caption{Expansion velocity of the bubbles for our simulations. The sound speed in the surrounding ISM is also indicated for several ISM temperatures.}
\label{FigBubbleVelocity}
\end{figure}

Unlike other studies \citep[\textit{e.g.}][]{Garcia-Segura1996a,Garcia-Segura1996b} where a double-shock structure appears in the simulation, our computations show only one shock (the wind termination shock) and no shock at the outer edge of the bubble in most cases. In our simulations, such a structure only appears in the very early evolution and in some cases in the WR state. This is due to the expansion velocity of the bubble through the ISM in our simulations, which is significantly subsonic for most of our models, except the most massive ones (above $60$-$85\, M_{\sun}$, see Fig.~\ref{FigBubbleVelocity}). An outer shock, able to significantly compress the ISM matter, thus can only develop for these very massive star models above approximately $60$-$85\, M_{\sun}$ or during the first few hundred thousand years of the star's lifetime. For stars with a lower mass and subsonic expansion, no outer shock can form and no shock-compressed shell can establish. Instead, ISM matter piled up by the stellar wind is simply pushed away by a pressure wave at approximately the sound speed.

Given the critical role of the bubble's expansion velocity with respect to the sound speed of the ISM, one may ask about the role of the temperature of the ISM, which determines its sound speed. The ISM temperature itself depends on the degree of ionisation. Taking ionisation effects into account in our simulations would be desirable, but is not feasible for as many models as we study here, owing to computational limitations. Looking at the literature, results from radiation-hydrodynamical models for single-star models that include ionisation effects show somewhat conflicting results: for a $40\,M_{\sun}$ star, \citet{Toala2011a} find a completely ionised ISM and swept-up shell ($\rho_\text{ISM} =100\,\text{cm}^{-3}$), whereas for a similar star \citet{Dwarkadas2013a} find that the swept up shell is only partly ionised ($\rho_\text{ISM} = 1\,\text{cm}^{-3}$). However, even in this latter case, the temperature of the swept-up shell is always higher than $1000\,\text{K}$.

Returning to our own study, we take from Fig.~\ref{FigBubbleVelocity} that even for such a low temperature, the expansion velocity of the bubble is subsonic for MS O-type stars with a mass lower than about $40\,M_{\sun}$. Nebulosities based on a compressed swept-up shell around such stars thus seem unlikely to exist in the framework of the hypothesis assumed in our computations (fully ionised and homogeneous ISM, single star). By contrast, for MS O-type stars with masses above about $40\,M_{\sun}$, such nebulosities may exist. As mentioned above, taking ionisation into account could change this picture, particularly during the first stages. The simulation of the formation of HII regions in star-forming region requires physical input that is not included in our simulations \citep[see \textit{e.g.}][]{Tremblin2012a,Tremblin2012b}. Most probably, more detailed simulations than presented in this work are required to explain the observed bubbles around young OB-stars by the GLIMPSE survey \citep{Churchwell2006a}.

The case is different for WR stars. Several WR stars exhibit ejecta nebulae around them, which are observable in visible wavelength \citep[H$\alpha$,][]{Stock2010a}. These ejecta nebulae are interpreted as resulting from the interaction of slow and dense matter ejected during a previous mass-loss episode, with the fast wind of the WR star. Most of the nebulae analysed by \citet{Stock2010a} are found around WN type WR stars (10 out of 13 in the Milky-Way sample). This is coherent with the picture that emerges from our simulations, showing that the dense region around the central star created during the RSG phase is progressively diluted during the WN phase, thereby decreasing its density, hence its observability. Moreover, in the framework of single-star evolution, the stars that have a RSG phase and then a WR phase finish their lives as WN stars \citep{Georgy2012b}. We thus expect most of the WR ejecta nebulae to be found around WN stars. The relatively small number of WN stars showing such a nebula \citep[$8\%$, according to][]{Stock2010a} is also understandable because the nebula is only visible close to the WR star (and thus only at the beginning of the WR stage), and also because several WN stars originate in more massive progenitors that completely avoid the RSG phase \citep{Georgy2012b}.

\subsection{Impact of the grid resolution}\label{SectResolution}

The simulations presented in this paper have a lower resolution than other recent work \citep[\textit{e.g.}][]{vanMarle2012a,Krause2013a}. The choice is deliberate, for it allows us to compute the ISM time evolution for a whole grid of stellar models, an impossible endeavour to tackle at much higher resolution owing to CPU time limitations. The price to pay is a restricted ability to catch the development of small-scale instabilities. However, since our aim is to describe the broad chemical features and bubble sizes, this limitation is not a severe one as we demonstrate in the following.

\begin{figure}
\begin{center}
\includegraphics[width=.5\textwidth]{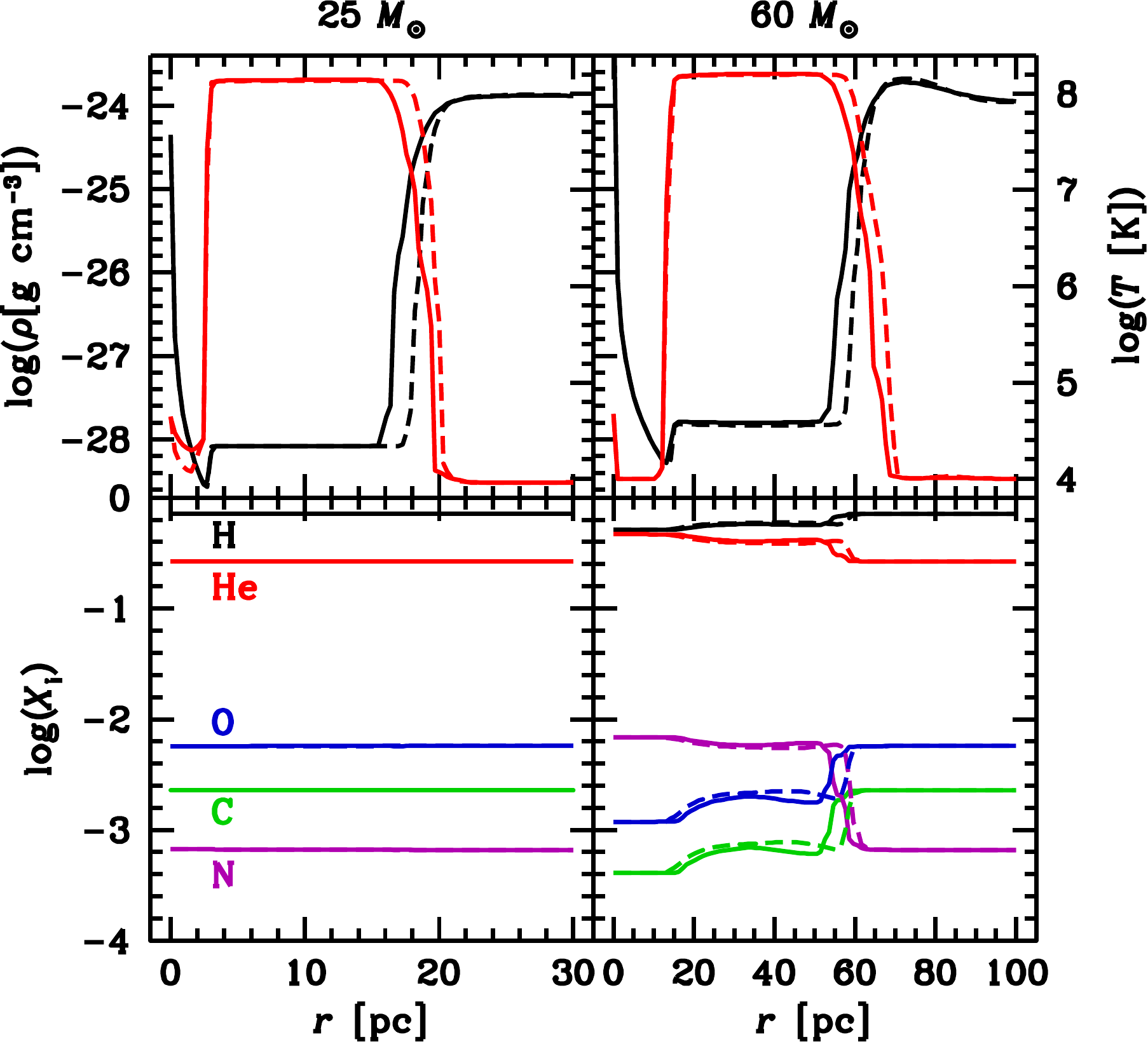}
\end{center}
\caption{\textit{Top panels:} Latitudinal averaged value of $\rho$ (black curves, left axis) and $T$ (red curves, right axis) as a function of the radius between the standard (solid line) and high (dashed line) resolution simulations. \textit{Bottom panels:} Averaged chemical mass fraction as a function of the radius. The left panels show the $25\,M_\sun$ at a time of $1.7\,\text{Myr}$, and the right panels the $60\,M_\sun$ at a time of $3\,\text{Myr}$.}
\label{FigResCompa}
\end{figure}

Doubling the spatial resolution of our grid, we recomputed four simulations: the $25\,M_\sun$ and $60\,M_\sun$ models, since they are representative of the evolution of the stellar winds studied here, against ISM densities of  $\log(\rho_\text{ISM} [\text{g}\cdot \text{cm}^{-3}]) = -24$ and  $\log(\rho_\text{ISM} [\text{g}\cdot \text{cm}^{-3}]) = -22$. The evolution of the $25\,M_\sun$ was followed up to $1.7\,\text{Myr}$, and the $60\,M_\sun$ up to $3.0\,\text{Myr}$. Results from the resolution study are compared in Fig.~\ref{FigResCompa} for the low-density ISM case. The high-density ISM case behave similarly. As can be seen, the bubble is slightly larger in the better resolved simulation. The contact interface is steeper, and its inner edge is shifted outward by about 9\%. The resolution effect is smaller on the outer shock. The steepening of the contact interface results because its  width in terms of grid cells remains about constant, independent of numerical resolution (about 3 to 5 cells for the high-resolution integrator used here). Its smaller spatial extension is crucial since the interface is the main cooling region of the bubble. Here, the density increases rapidly from the low values in the bubble interior towards the high value found in the dense region of cold ISM material \textit{and}, at the same time, the temperature decreases from the very high value of the shocked wind. The intermediate temperature and density values are the most favourable for radiative cooling. Thus, in summary, higher numerical resolution results in a thinner contact interface and reduced radiative losses, thus more energy remains to drive the bubble expansion and the bubble evolves slightly faster than in the simulation with lower numerical resolution (see also the blue and red curves in Fig.~\ref{FigScaling}). This effect is well known and described, for instance, in \citet{Folini1995a}, \citet{Walder1996a}, or \citet{Parkin2010a}.

Even if the size of the bubble shows some sensitivity to the resolution, results are not too different with regard to latitudinal averages of the density, temperature, and chemical composition (see Fig.~\ref{FigResCompa}). We see that except for the small shift produced by the slightly different size of the bubble at any given time, the averaged quantities we consider in this work are remarkably similar.

From the above results and considerations we conclude that resolution effects are not very important for our main conclusions, which are targeted at large-scale features. This is not to say that much higher resolutions or 3D simulation, both potentially allowing for more and different instabilities, would not add to the picture arising from the present study.

\subsection{Non-self-similar evolution}\label{Sec:Results_Selfsimilarity}

\begin{figure*}
\begin{center}
\includegraphics[width=.48\textwidth]{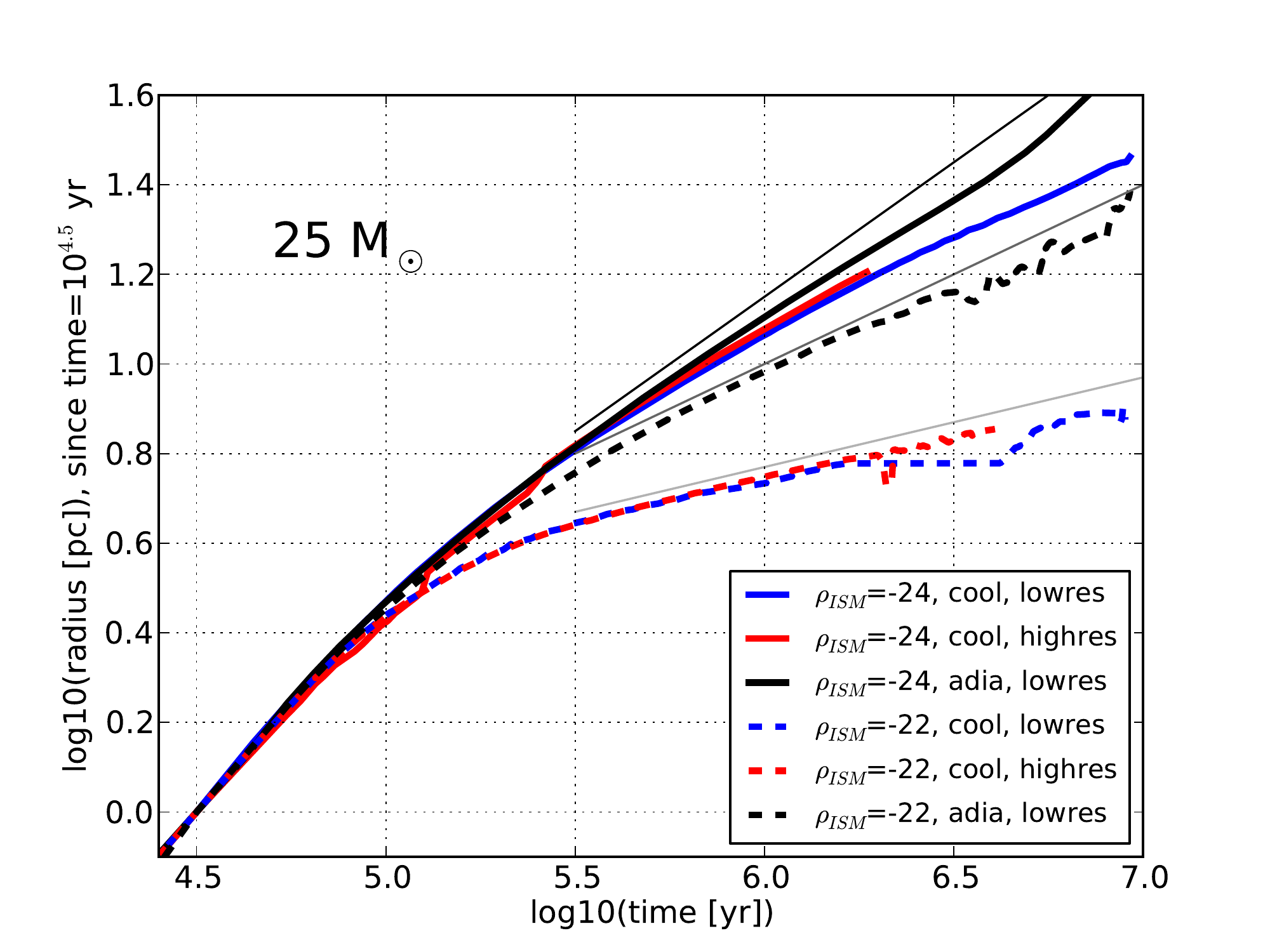}\hfill\includegraphics[width=.48\textwidth]{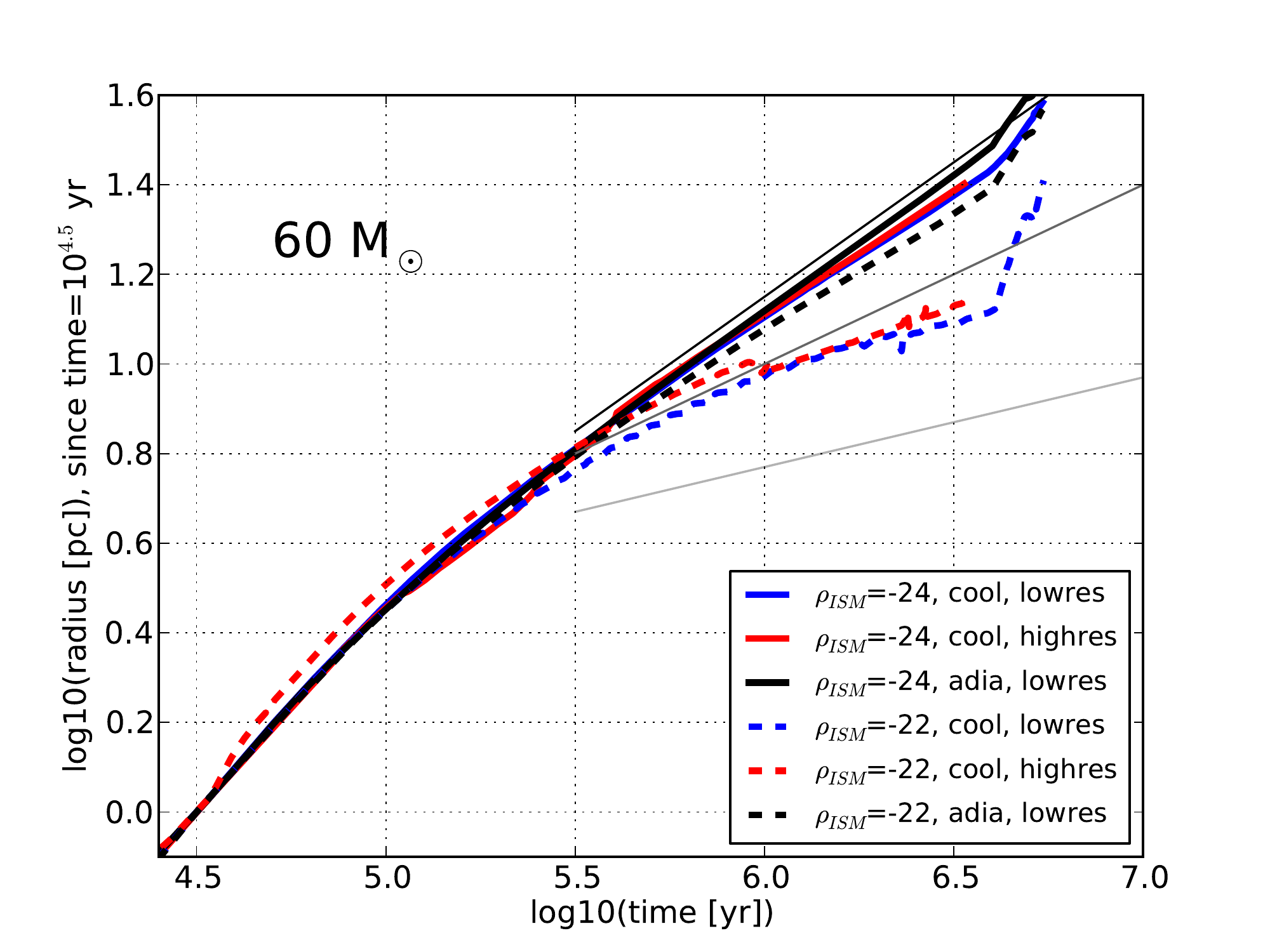}
\end{center}
\caption{\textit{Left panel:} Evolution of the radius of the bubble as a function of the time for various $25\,M_\sun$ models. All models are normalised for the time of $\log(t\,\text{[yr]}) = 4.5$. The extension of the bubble at this time is about $0.75\,\text{pc}$, with small deviations for the different models. \textit{Right panel:} Same for the $60\,M_\sun$ models. Here the extension of the bubble for the normalisation time is about $1.7\,\text{pc}$. Thin lines indicate slopes of $3/5$, $2/5$, and $1/5$.}
\label{FigScaling}
\end{figure*}

\citet{Weaver1977a} have shown that a bubble evolves in a self-similar way under many circumstances. For instance, for a purely adiabatic evolution (polytrophic EOS with $\gamma = 5/3$), a constant wind power, $L_\text{wind}$, a constant density of the ISM $\rho_\text{ISM}$, and a negligible pressure of the ISM, the position of the shock towards the ISM (index s) and the position of the contact discontinuity (index c) evolve as $r_{s, c} = \alpha \left(L_\text{wind}/\rho_\text{ISM}\right)^{1/5}t^{3/5}$, where $\alpha$ is a parameter of order unity. The velocities of these waves scale as $v_{s, c}\propto t^{-2/5}$. However, as also discussed by \citet{Weaver1977a}, the bubble does not always evolve in a self-similar way, in particular when radiative cooling affects its dynamics or when the self-similarity is broken by the presence of a substantial pressure in the ISM.

Figure~\ref{FigBubbleVelocity} and Table~\ref{TabRhoZ} indicate that in the presented simulations a self-similar evolution is the exception. This is confirmed by a closer analysis, as summarised in Fig.~\ref{FigScaling}. Indeed, this can be expected, since radiative cooling is included in our models, the wind power is not constant over the evolution (Fig.~\ref{FigWindPower}), and the pressure of the ISM (in particular in the models with a dense ISM) cannot be neglected. To understand the behaviour of our models in this context, we computed additional models: some with an increased resolution (factor of 2 in each direction, see Sect.~\ref{SectResolution}), and some without any radiative cooling (adiabatic case).

In a first phase, present in all our simulations, the size of the bubble increases linearly, $r_{s, c} \propto t$, approximately until $\log(t\,\text{[yr]}) = 5$. The duration of this linear phase depends somewhat on the stellar mass and ISM density. The reason for the existence of this linear phase is --similar to the case of a supernova remnant \citep[see \textit{e.g.}][]{Chevalier1982a}-- that sufficient ISM material has to be collected in front of the snow plough before the interaction zone slows down and adopts self-similar behaviour.
  
In a next phase, we find clear differences between the adiabatic models (black in Fig.~\ref{FigScaling}) and the models including radiative cooling and using different numerical resolutions (blue and red lines in Fig.~\ref{FigScaling}). While some models show self-similar evolution of the form  $r_{s, c} \propto t^{3/5}$, others show a shallower dependence on time. A clear $3/5$ law is seen for the adiabatic  $60\,M_{\sun}$ model against $\log(\rho_\text{ISM})=-24$. This model essentially meets the underlying assumptions (see above) of the self-similarity law. The self-similar phase lasts until the sharp rise in wind power close to the end of the star's life (Fig.~\ref{FigWindPower}). Small deviations from the self-similar evolution we ascribe to the non-constant wind-power even during the MS evolution.

Reduction of the wind power tends to reduce the slope below $3/5$. The $25\,M_{\sun}$ adiabatic model with $\log(\rho_\text{ISM})=-24$ already deviates towards a power law coefficient that is clearly smaller than $3/5$. The deviation starts when the pressure of the ISM becomes similar to the pressure of the shocked wind. Similarly, a denser ISM ($\log(\rho_\text{ISM})= -22$, dashed lines in Fig.~\ref{FigScaling}) also leads to an earlier and more pronounced deviation from $3/5$. For the $25\,M_{\sun}$ model, the slope is close to $2/5$.

Radiative cooling also tends to reduce the slope of the power law evolution, as part of the  wind energy is radiated and cannot be used to drive the bubble. While the radiatively cooling  $60\,M_{\sun}$ model against $\log(\rho_\text{ISM})= -24$ still has a slope close to $3/5$, the $60\,M_{\sun}$ model against $\log(\rho_\text{ISM})= -22$ has a shallower slope close to $2/5$, and the $25\,M_{\sun}$ model against $\log(\rho_\text{ISM})= -22$ has a much shallower slope of only about $1/5$. The much more pronounced dependency on the ISM density, as compared to the adiabatic case, is expected since cooling is proportional to density squared. Nevertheless, the ISM density does not translate directly into slope changes, because the main cooling region of the bubble is the contact interface where temperature and density take intermediate values between the hot bubble and the cold ISM. This is detailed in Sect.~\ref{SectResolution} and in the literature~\citep{Folini1995a, Walder1996a, Parkin2010a}, along with an explanation of why higher spatial resolution results in reduced radiative losses and a slightly faster bubble expansion, i.e. a slightly steeper slope for the high-resolution simulations, as is apparent in Fig.~\ref{FigScaling} (red lines are steeper than blue lines).

We conclude that (1) bubbles do not generally evolve self-similarly over their entire lifetimes and (2) that changes in the ISM density or the metallicity of the star cannot be taken into account by a simple analytical formula. Instead, the values given in Table~\ref{TabRhoZ} mirror these differences much better.

\section{Discussion and conclusions}\label{SecConclu}

\begin{figure}
\begin{center}
\includegraphics[width=.5\textwidth]{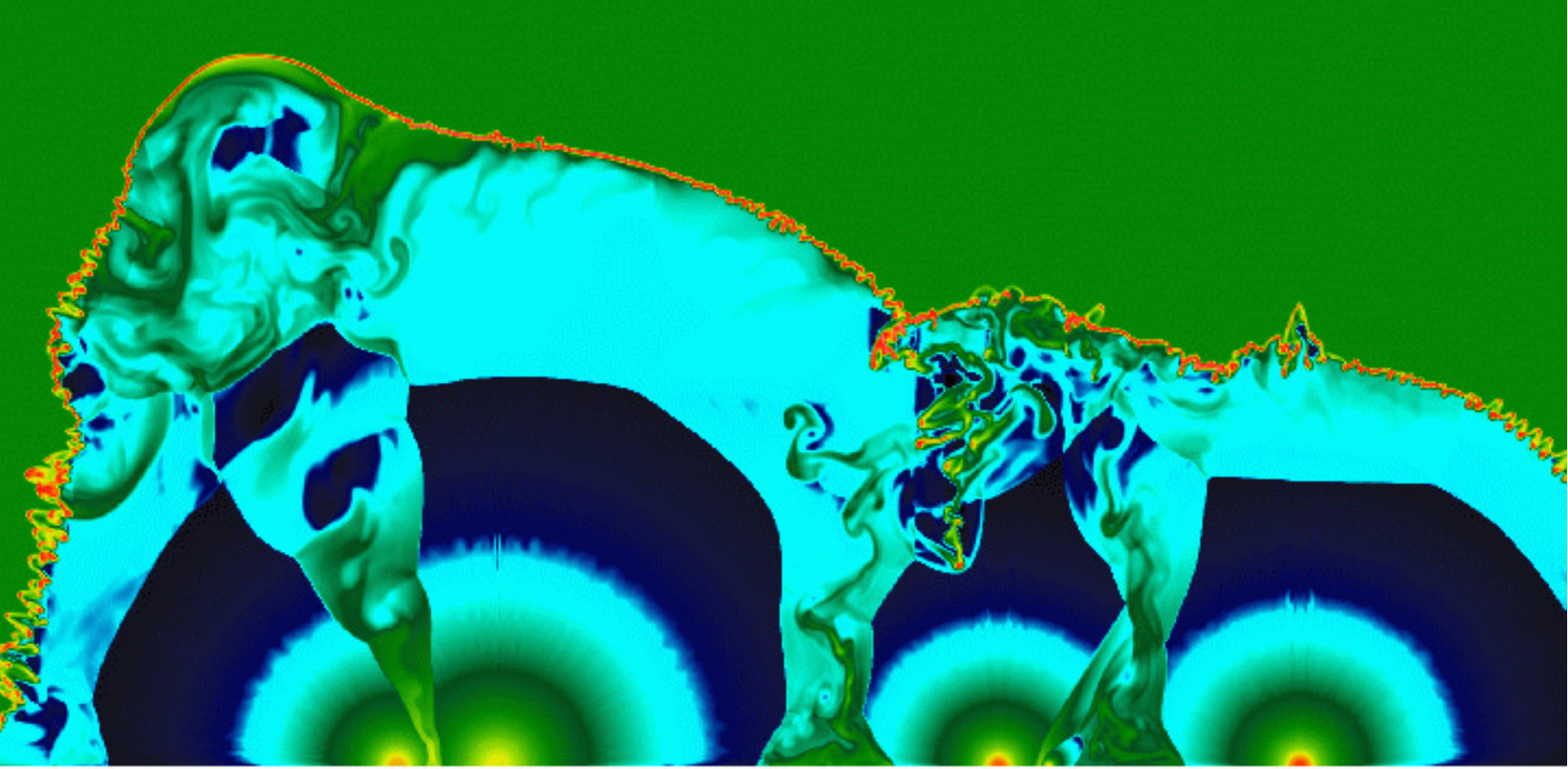}
\end{center}
\caption{Axisymmetric simulation of a mini-star cluster of 5 stars shown in density (blue: 0.01/cm$^3$, green 1/cm$^3$, red 10/cm$^3$, white 1000/cm$^3$). Thin shell instabilities develop in regions of interactions of different winds and of wind-interstellar medium.}
\label{FigStellarSystem}
\end{figure}

 Returning once more to the complex flow shown in Fig~\ref{FigStellarSystem}, we briefly want to highlight the potential impact of such flow conditions for high-energy astrophysics. Future studies in this direction are envisaged, profiting from our enlarged toolbox, especially the elaborate implementation of stellar systems as described in Sect. 2. Powerful stellar winds of young massive stars and core-collapse supernova explosions with strong shock waves can convert a sizeable part of the kinetic energy release into fluctuating magnetic fields and relativistic particles. The star-forming regions and compact young star cluster are considered as favourable sites for energetic particle acceleration and could be seen as bright sources of non-thermal emission with the upcoming high-energy instruments, either at hard X-rays or at TeV gamma-rays by the Cerenkov Telescope Array (CTA).

Rich associations of OB-stars, particularly Cygnus OB, have recently been detected at gamma-rays. The Fermi Large Area Telescope has detected $1-100\,\mathrm{GeV}$ photon emission from a 50-parsec-wide cocoon-like structure in the active Cygnus X star-forming region \citep[][]{Ackermann2011a}. The authors proposed that the cocoon is filled with  freshly accelerated cosmic rays that flood the cavities carved by the stellar winds  from young stellar clusters. The gamma-ray luminosity estimated is about $10^{35}\,\mathrm{erg/s}$, which is a small fraction of the kinetic power of the stellar winds. However, to produce the observed gamma-ray luminosity, one should maintain much greater power in cosmic rays, and this requires an efficient conversion of the kinetic power of the stellar winds and supernovae. An extended source of very high energy emission was also detected with the High Energy Stereoscopic System (H.E.S.S.) from Cyg OB2 region \citep[][]{Aharonian2002a} and a very massive young compact cluster, Westerlund I \citep[][]{Abramowski2012b}. Starburst galaxies NGC 253, M82, and some others \citep[see for a recent review][]{Ohm2012a} have demonstrated the high-energy emission spectra that are harder than that of the Milky Way or M31, where the global diffuse TeV regime emission has not been reported so far.

Complex supersonic flows carrying magnetic fluctuations of a broad dynamical range of scales can efficiently interact with relativistic particles. The interaction results in particle acceleration, which in turn modifies the plasma flows and affects specific mechanisms of magnetic field amplification in the vicinity of collisionless shocks \citep{vanMarle2012b}. Relativistic particles are subject to Fermi acceleration at strong astrophysical collisionless shocks \citep[see for a review][]{Blandford1987a}. Ensembles of multiple shocks accompanied with large-scale MHD motions are very efficient particle accelerators \citep[\textit{e.g.}][]{Bykov2001a,Bykov2013a,Parizot2004a,Ferrand2010a}, and they can be the sources of the
galactic cosmic rays accelerated beyond the spectral knee \citep[\textit{e.g.}][]{Bykov2001b}.

In this paper, we present hydrodynamical simulations of the time evolution of the CSM around massive stars ($15$ to $120\, M_{\sun}$), based on the recent grid at solar metallicity provided by \citet{Ekstrom2012a}. We show the differences between various evolutionary scenarios, by extracting from our data various averaged quantities as a function of (stellar evolution-) time and/or the distance to the central star: radius of the bubble, position of the wind termination shock, density, temperature, and chemical composition of the gas. The differences found, which cannot be overcome by simple scaling laws, demonstrate the need for grids of models as a complement to very detailed (and computationally expensive) studies of one particular evolutionary track.

For the range of models considered, we find bubble radii right before the SN explosion between roughly 10 and 100 pc. Much smaller bubbles are obtained for higher ISM densities (factor 3 for ten times higher density) and/or lower metallicities of the massive star. Average densities within the bubble are generally (much) lower than ISM densities, except very close to the star. The chemical composition in the bubble can be very inhomogeneous and very different from the ISM composition, owing to the progressive modification of the chemical composition of the star. This finding is in line with observations of the WR nebula NGC 6888. 

Concerning, more generally, the (non-) existence of observable nebulae around massive stars, our models allow for WR ejection nebulae (WR wind against RSG wind) around young WN type WR stars, in line with observations. On the scale of the entire bubble, our models suggest that the wind blown bubble is confined by an (observable) outer shock wave only for the most massive stars and then only during their WR phase. During MS, a shock at the outer rim of the bubble forms only at the very early phases or for the most massive models. Later on, no such shock is present, because the bubble expands subsonically. A shock front may, however, still develop from the action of the photoionisation front \citep{Toala2011a} or if hydrogen can recombine and the gas can cool to very low temperatures \citep{Dwarkadas2013a}. A complete view of the formation of nebulosity around MS massive stars thus requires taking physical effects into account that are not yet included in our simulations.

Leaving the concept of single stars in favour of stellar clusters, wind collision zones very likely alter the above results: mixing of chemical species is likely more efficient, and observable signatures more pronounced.

At the end of their lives, the models presented in this work explode as a SN. The question of how these explosions disperse the chemical species expanding through the previously formed bubble is important in the matter of the chemical evolution of the galaxies. This point will be studied in a forthcoming paper.

%=================================================================================
% ACKNOWLEDGEMENTS
%=================================================================================
\begin{acknowledgements}
The authors thank the anonymous referee for his pertinent comments that have significantly improved the quality of the present work. The authors thank the P\^ole Scientifique de Mod\'elisation Num\'erique in Lyon for the access to their computing facilities. They are also grateful to R. Hirschi, for allowing them to use the Shyne cluster to perform part of the simulations presented in this paper. CG acknowledges support from the European Research Council under the European Union's Seventh Framework Programme (FP/2007-2013) / ERC Grant Agreement n. 306901, as well as from the Swiss National Science Foundation. CG, RW, and DF acknowledge support from the Programme National de Physique Stellaire. AB was supported by Russian Academy of Sciences  Presidium and OFN-17 programme.
\end{acknowledgements}

% BIBLIOGRAPHIE %%%%%%%%%%%%%%%%%%%%%%%%%%%%%%
\bibliographystyle{aa}
\bibliography{MyBiblio}
%%%%%%%%%%%%%%%%%%%%%%%%%%%%%%%%%%%%%%%%

\end{document}